\documentclass[reprint, prx]{revtex4-2}

\usepackage{graphicx}
\usepackage{subfig}
\usepackage{amsmath}
\usepackage{natbib}
\usepackage{url}
\usepackage{lipsum}
\usepackage{xcolor}

\usepackage{apptools}
\AtAppendix{}

\makeatletter
\newcommand*{\balancecolsandclearpage}{%
  \close@column@grid
  \clearpage
  \twocolumngrid
}
\makeatother

\begin{document}

\title{Congruity of genomic and epidemiological data in modeling of local cholera outbreaks}

\author{Mateusz Wilinski$^1$, Lauren Castro$^2$, Jeffrey Keithley$^{2, 3}$, Carrie Manore$^1$, Josefina Campos$^4$, Ethan Romero-Severson$^1$, Daryl Domman$^{5*}$, Andrey Y. Lokhov$^1$}
\thanks{ddomman@salud.unm.edu, lokhov@lanl.gov}
\def\thefootnote{\arabic{footnote}}
\affiliation{$^1$ Theoretical Division, Los Alamos National Laboratory, Los Alamos, NM USA}
\affiliation{$^2$ Analytics, Intelligence and Technology Division, Los Alamos National Laboratory, Los Alamos, NM USA}
\affiliation{$^3$ Department of Computer Science, University of Iowa, Iowa City, IA, USA}
\affiliation{$^4$ UO Centro Nacional de Genomica y Bioinformática, ANLIS “Dr. Carlos G. Malbrán”, Buenos Aires, Argentina}
\affiliation{$^5$ Center for Global Health, Department of Internal Medicine, University of New Mexico Heath Sciences Center, Albuquerque, NM USA}

\begin{abstract}
Cholera continues to be a global health threat. Understanding how cholera spreads between locations is fundamental to the rational, evidence-based design of intervention and control efforts. Traditionally, cholera transmission models have utilized cholera case count data. More recently, whole genome sequence data has qualitatively described cholera transmission. Integrating these data streams may provide much more accurate models of cholera spread, however no systematic analyses have been performed so far to compare traditional case-count models to the phylodynamic models from genomic data for cholera transmission. Here, we use high-fidelity case count and whole genome sequencing data from the 1991-1998 cholera epidemic in Argentina to directly compare the epidemiological model parameters estimated from these two data sources. We find that phylodynamic methods applied to cholera genomics data provide comparable estimates that are in line with established methods. Our methodology represents a critical step in building a framework for integrating case-count and genomic data sources for cholera epidemiology and other bacterial pathogens.
\end{abstract}

\maketitle

\section*{Introduction}
Cholera is a major public health threat with an estimated 4 million cases a year and over 150,000 deaths annually \cite{ali_updated_2015}. 
Cholera is an acute diarrheal disease caused by toxigenic \textit{Vibrio cholerae}, and is transmitted though the fecal-oral route from contaminated food or water.
Currently, the burden of disease is primarily in sub-Saharan Africa and South Asia, in  vulnerable populations with a lack of access to clean drinking water and sanitation \cite{clemens}.   
Of note, some of the largest cholera epidemics have occurred since the 1990s. 
Case counts over 1 million were documented for the 1991-1998 epidemic in Latin America \cite{kumate_cholera_1998}, over 800,000 cases in Haiti from 2010-2020 \cite{noauthor_cholera_2021}, and over 2.5 million cases thus far in the ongoing epidemic in Yemen that began in 2016 \cite{noauthor_cholera_nodate}.

Understanding how cholera is transmitted within and across populations is paramount to the rational design and implementation of control efforts. One of the difficulties in traditional modeling of cholera outbreaks with case-count data alone is that the inference results strongly depend on the quality of the reporting procedures, while detailed properties of statistical counting noise are often unknown and can not be easily estimated.
Furthermore, accurate case counting is hampered by the large heterogeneity in disease presentation which ranges from asymptomatic to severe cholera \cite{clemens}.
Genetic sequence data offers another data source on transmission dynamics that could help develop a new generation of complex, but well constrained cholera models.
This is possible due to the fact that epidemiological processes such as transmission and migration of infection leave a trace in pathogen genetic sequence data by changing the underlying infection genealogy of a sampled set of sequences \cite{volz_viral_2013, rosenberg_genealogical_2002, grenfell_unifying_2004}.
For example, compared to a stable population, in a population experiencing rapid growth, two randomly selected people will, on average, share a common ancestor in the distant past \cite{wakeley_coalescent_2009}, and therefore be separated by a larger number of mutations.
Likewise, in otherwise isolated populations, migration links pathogens though a network of common descent \cite{nath_coalescent_1993} leading to a distinctive pattern of interdigitated sequences in a phylogeny.
Therefore, pathogen genetic sequence data has generally the potential to inform epidemiological parameters such as transmission, migration, and mixing rates.
Previous work have used \textit{Vibrio cholerae} sequence data to describe the broad, qualitative flow of cholera at the global scale that can capture broad trends but not explicit details of the transmission process \cite{weill_genomic_2017, domman_integrated_2017, weill_genomic_2019, mutreja_evidence_2011, didelot_role_2015}. 
However, the main challenge in the effort to integrate genomic data into cholera transmission models is the lack of evidence that genomic data are informative of cholera transmission processes at the local scale, i.e. that there is sufficient genetic diversity in a single-source, local cholera outbreak to estimate transmission parameters.

Case counts and the genetic sequence data can be thought of as being two independent observers of the transmission dynamics of cholera. In this paper, we utilize a rich data-set of both case count data as well as a large collection of genomic sequencing data from Argentina \cite{dorman2020genomics} to model the spread of cholera and specifically address the role of local migration as a driving factor in cholera transmission. Our goal is to build a meta-population model with a minimal number of assumptions which accounts for migration, and understand if the two observers agree in their predictions. The bridging and common constraints on both sources of data will be achieved using yet another source of independent data such as high-fidelity estimation of migration flows.

Our modeling choices are aimed at simplicity and driven by the overall goal of checking the consistency between these data sources. For instance, we deliberately take an agnostic stance on the open questions related to the role of the environment or details of bacterial dynamics: while several previous studies explicitly included the environmental compartment \cite{magny2005modeling, hartley2005hyperinfectivity, king2008inapparent, kolaye2019mathematical, meszaros2020direct} leading to a larger number of model parameters, we choose to model cholera dynamics as an \emph{effective} transmission process which includes a periodic functional dependence on the seasonality, similarly to the approach of \cite{peak2018prolonging}. To account for discreteness in observed cases, we propose a novel sampling model which relates the continuous model with the discrete observed case counts.

Most of key model parameters will be directly inferred from case counts and migration data. Further, a subset of most important parameters such as transmission amplitudes and fraction of asymptomatic infections are independently inferred from the genomic sequence data, and compared to models inferred from other data sources within their uncertainties. In particular, we don't make any \emph{a priori} quantitative assumptions on the fraction of asymptomatic infections, in previous studies ranging from 1\% to more than 90\% of the population  \cite{mondiale2017weekly, rafique2016transmission, mccormack1969community}. Instead, we keep this important model parameter free, infer its values from data under different settings, and discuss the sensitivity of this parameter to various modeling assumptions. We also provide a series of careful sensitivity studies that study the stability of the inferred parameters related to all of our modeling assumptions.

In this paper, we present initial evidence that phylodynamic methods can be used to study cholera outbreaks at a regional level and that they produce parameter estimates that are consistent with established methods. Our approach provides a common methodology for an early analysis of the model viability in the context of joint inference from different data sources. Given the complementary view offered by independent data sources, we anticipate that the analysis presented in this paper will find a widespread use in building joint hybrid epidemiological and genetic models which could help verify the main modeling assumptions.

\section*{Results}
\begin{figure*}
    \centering
    \includegraphics[width=0.2814\textwidth]{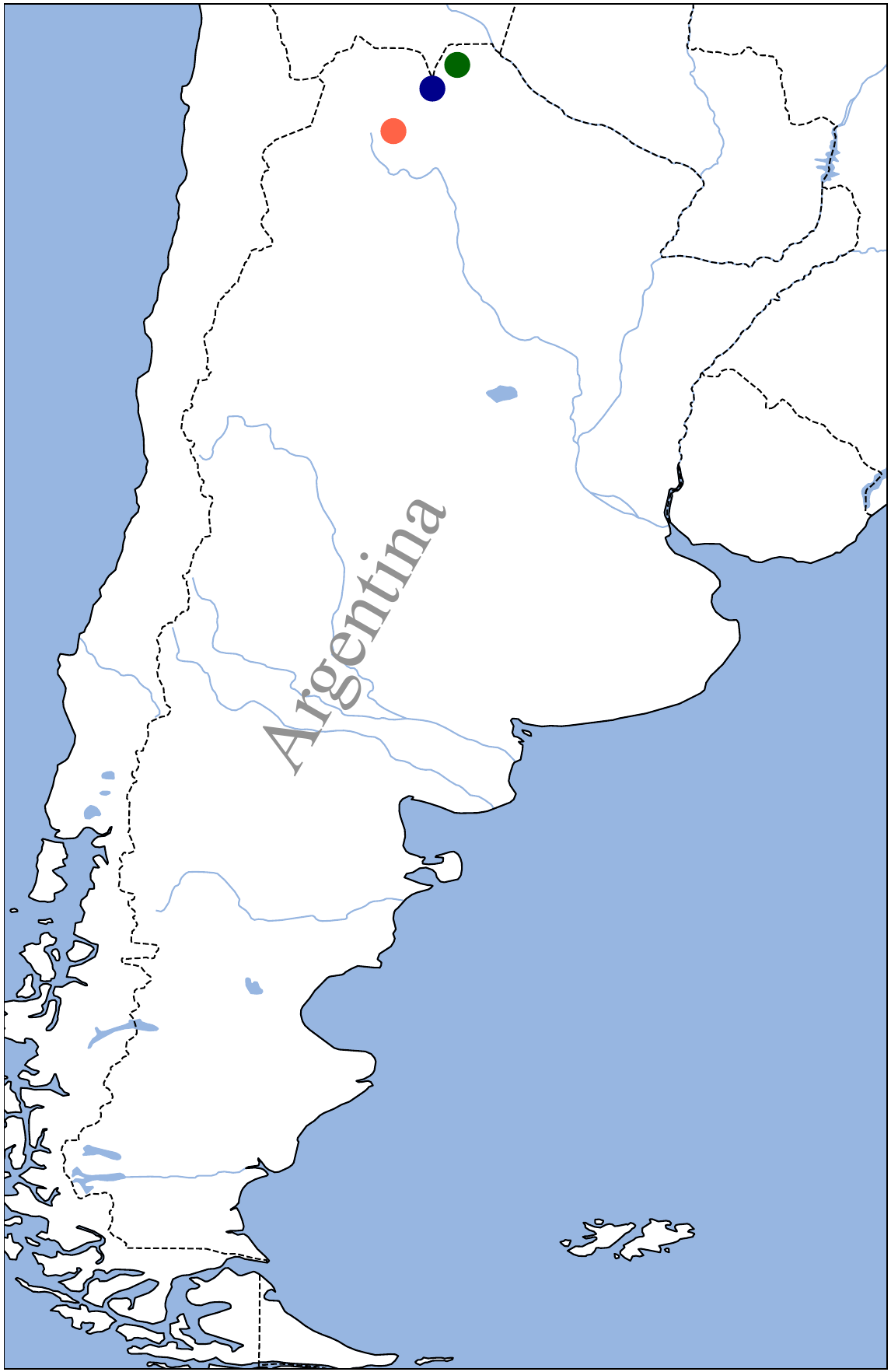}
    \includegraphics[width=0.5\textwidth]{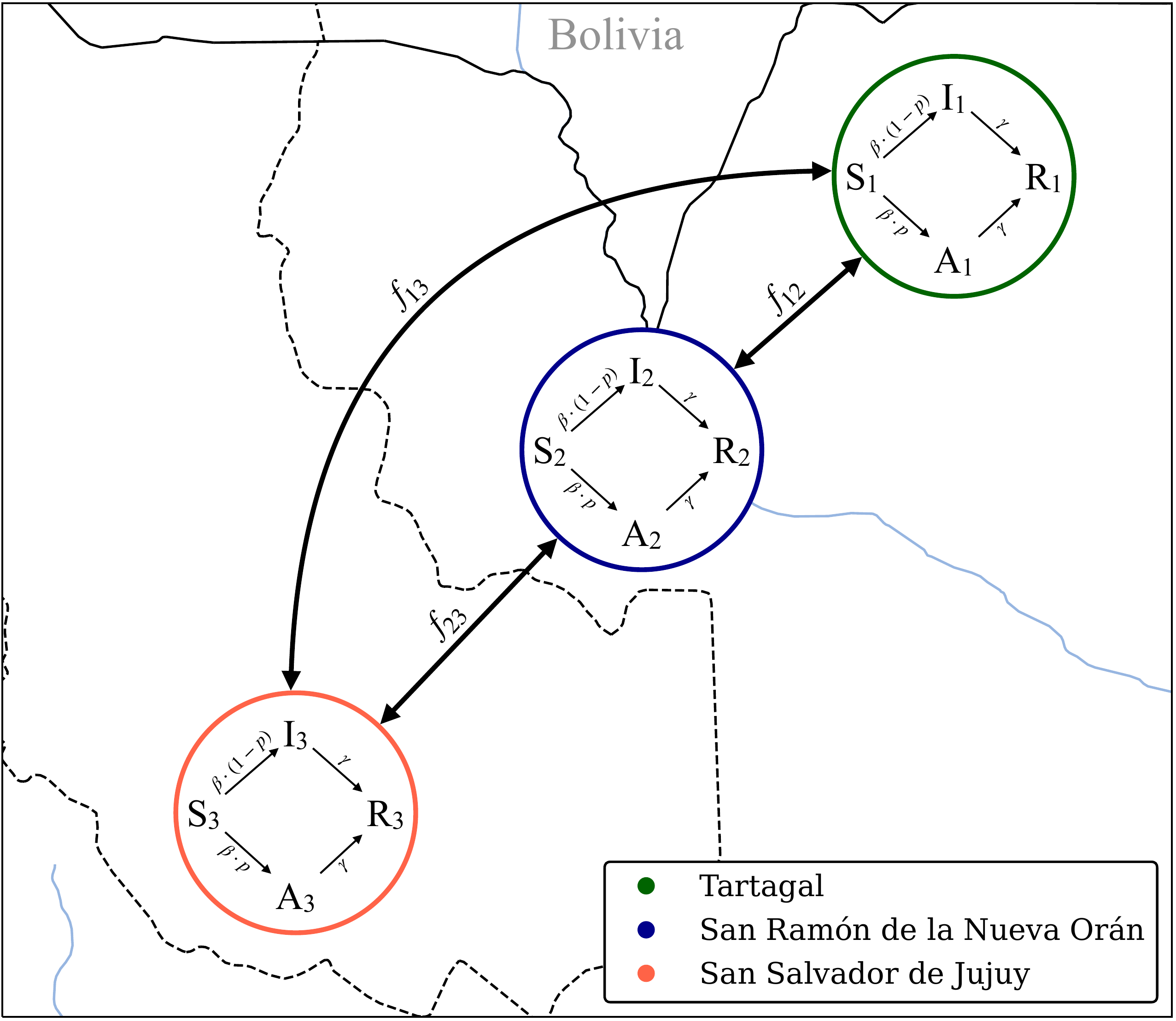}
    \caption{\label{fig:map} Meta-population model of the cholera transmission dynamics interlayed with a map of the northern Argentina. Left: Three cities considered in our focused study are marked as follows:
    Tartagal -- green circle, San Ramón de la Nueva Orán (in what follows, referred to as Oran) -- blue circle -- and San Salvador de Jujuy (in what follows, referred to as Jujuy) -- orange circle. Right:
    The dynamics inside each city population is modelled using the Susceptible-Infected-Asymptomatic-Recovered (SIAR) model with seasonality modulated infection rate $\beta(t)$, recovery rate $\gamma$, and the parameter $p$ representing the fraction of asymptomatic cases under the infection process. The amplitude of the infection rate is $\beta_s$ for cities with smaller population (Tartagal and Oran), and $\beta_l$ for a larger city (Jujuy).
    Black arrows represent the migration flows of susceptible and asymptomatic individuals between cities, proportional to the flow rates $f_{ij}$ for migration between locations $i$ and $j$.
    A more detailed description of the model is provided in the \emph{Materials and Methods}, as well as in the \emph{Supplementary Materials, section} \ref{app:model}.}
\end{figure*}

{\bf Integrated data from case counts, genomics, and transportation data.}
Cholera was first reported in Argentina in 1992, and subsequent cholera cases were reported until 1998 \cite{report1992, report1993, report1994, report1995, report1997}.
Out of the total 4,281 cases reported, over 3,500 \textit{Vibrio cholerae} isolates were stored at INEI-ANLIS ``Dr. Carlos G. Malbrán'', the national reference laboratory for Argentina, and a representative sub-sample of 532 of these isolates were previously whole-genome sequenced \cite{dorman2020genomics}, see \emph{Supplementary Materials, section} \ref{app:samples} for more details.
We sought to determine if there was agreement between epidemiological and genomic data.  
First, we pre-processed the data set by removing cities with insufficient genomic samples (less than 40 sequences).
This left us with three target cities: Tartagal, San Ramón de la Nueva Orán (both in Salta province) and San Salvador de Jujuy (in Jujuy province) located within in the Northwest of Argentina (see Fig. \ref{fig:map}).
Initial reports in 1992 indicated that cholera was first introduced into Argentina via this region from Bolivia, leading to a large outbreak from 1992-1993  \cite{mazzafero1995epidemic}.

In addition to the epidemiological and sequence data, we used publicly available data on domestic travel to estimate the movement of population between these three cities during the study period.
Focusing on the two primary means of transportation, flights and buses, we were able to estimate the typical number of people travelling daily between the selected cities (for details see \emph{Materials and Methods} and \emph{Supplementary Materials, section} \ref{app:fij}).

{\bf Modeling assumptions.} We modeled the cholera transmission dynamics using a system of ordinary differential equations (ODEs) where the population is split into compartments representing individuals in different states of infection.
Typical cholera models are a system of ODEs representing a modification of the classical Susceptible-Infected-Recovered (SIR) type model \cite{anderson1992infectious} with varying degrees of complexity (see \cite{magny2005modeling, hartley2005hyperinfectivity, king2008inapparent, neilan2010modeling, peak2018prolonging, kolaye2019mathematical, meszaros2020direct}).
Here, we present a new, simple ODE cholera model that significantly advances estimation of key epidemiological parameters in two ways: (i) we focus on a minimalist representation which allows us to reliably infer model parameters from a limited amount of data while introducing the least amount of assumptions; and (ii) we use a meta-population structure to leverage the spatial knowledge on reported cases and travel patterns that can represent the major spreading mechanism. 
We also do not consider re-infection in our models of localized outbreaks, as protective immunity against cholera has been estimated to last at least 3 years \cite{clemens}. 
Prior to formally introducing our dynamic model, we discuss the main modeling assumptions behind our approach.   

Many cholera models in the literature include an environmental compartment \cite{magny2005modeling, hartley2005hyperinfectivity, king2008inapparent, kolaye2019mathematical, meszaros2020direct}. 
Such an environmental compartment is typically introduced to explicitly model the transmission of infection through a water source, and additionally describes the evolution of bacteria in a water source with a temperature-dependent dynamics.
From the fitting perspective, an environmental component may have a benefit to help the multi-year epidemic outbreak (see the span of observed cases in the \emph{Supplementary Materials}, Fig. \ref{fig:samples}) survive the period of cool temperatures when number of cases drop significantly, and re-occur when the temperature rises.
During our initial model exploration, we tested an extension of our model that included an environmental compartment, finding that its inclusion did not improve the quality of the fit, while at the same time it introduced additional parameters that needed to be inferred from data.
For this reason and in order to keep the number of model parameters small, we do not explicitly include the environmental compartment in our model.
Instead, the seasonal component of cholera transmission, well documented in \cite{faruque2005seasonal, emch2008seasonality, daisy2020developing}, is included in a direct transmission parameter of our model. This direct contact parameter is an \emph{effective} parameter describing the spread of cholera which includes all potential transmission channels, similarly to an approach used in \cite{peak2018prolonging}. 
Previously, seasonally-modulated transmission parameters was suggested in a fully theoretical framework in \cite{emvudu2013mathematical}, but it was not applied to empirical data.

Our second key assumption is related to the presence of an asymptomatic population, which does not display any strong symptoms, but nevertheless contributes to the infection spread via migration. This population is not directly observed, but contributes to the cholera dynamics via a dedicated compartment $A$. The associated parameter $p$ describes the fraction of infected population which falls into the $A$ compartment upon infection, while the rest of the population falls into the $I$ compartment which describes the symptomatic population.
In spite of the general agreement that a significant number of individuals infected by cholera display no apparent symptoms and play a crucial role in spreading the disease between different locations, the literature is not conclusive about the proportion of asymptomatic carriers.
For instance, \cite{mondiale2017weekly} suggests that between 1\% and 25\% of infected cases are asymptomatic; \cite{rafique2016transmission} estimates $p$ closer to 50\%; and according to \cite{mccormack1969community}, the asymptomatic population represents the majority of cases.
Therefore, we treat $p$ as one of the key free parameters in our model which will need to be inferred from data. We further assume that the transmission dynamics between cities through migration is mediated by the asymptomatic carriers only.
This assumption is incorporated into the meta-population model through a migration term which is proportional to the number of asymptomatic individuals in the population, and is appropriately normalized so that the city populations do not change. Additionally, these migration terms are informed by independently estimated travel rates, as we discuss below. These migration terms link epidemic trajectories in different cities and thus facilitate the identification of the parameter $p$ related to the fraction of asymptomatic cases upon infection.

{\bf Meta-population model.}
Here, we formally summarize our dynamic meta-population model.
The cholera dynamics in each of the three cities is described using a homogeneous SIR-like SIAR model, linked by a flow of asymptomatic infected individuals between cities.
More precisely, we divide the population of each city into four compartments: susceptible (S), infected symptomatic (I), infected asymptomatic (A) and recovered (R).
The migration mechanism allows for a mixing of different city populations with two conditions: (i) symptomatic infected individuals do not move between locations; (ii) traveling on average does not change the population of each city.
A schematic representation of the structure of the meta-population model and the details of the single-city model are shown in Fig. \ref{fig:map}.
The exact system of ordinary differential equations used to build the model is described in detail in the \emph{Supplementary Materials, section} \ref{app:model}.

The model contains a total of 10 epidemiological, travel, and demographic parameters.
The epidemiological parameters are the recovery rate $\gamma$, which represents the inverse of expected days to recovery, the parameter $p$, which represents the fraction of asymptomatic cases emerging upon infection, and the infection rate $\beta(t)$ modulated by a time-dependent function reflecting the seasonal changes.
Importantly, while the seasonality itself is assumed to be the same for all three geographically close cities, the amplitudes $\beta_s$ and $\beta_l$, respectively representing the smaller population cities Tartagal and San Ramón de la Nueva Orán, and the larger population San Salvador de Jujuy, may be different.
The transmission amplitudes $\beta_s$ and $\beta_l$ can \emph{a priori} take different values, for instance due to an expectation of a better infrastructure in larger cities, potentially leading to access of higher quality health care and resulting in lower infection rates.
Additionally, the model parameters include the initial demographic structure of all four compartments in each location.
Finally, the model contains a set of migration parameters $f_{ij}$ describing the flow of people between the cities. The procedure for estimating the migration parameters is given in the \emph{Supplementary Materials, section} \ref{app:fij}.
A more detailed description of the fixed and free parameters is included in the \emph{Materials and Methods}, as well as in the \emph{Supplementary Materials, section} \ref{app:model}.

{\bf Estimation of model parameters from case count data.}
Using a least-squares-based estimator minimizing the error between the case counts data and model predictions (see \emph{Materials and Methods} for more details), we estimated the transmission rates, seasonality parameters, initial conditions, and asymptomatic fraction $p$. One of the main challenges faced by the fitting of our continuous meta-population SIAR model was the fact that the case counts are not reported continuously in the data set, but instead appear as discrete peaks at certain sampling days. To address this reporting delay challenge, we proposed a sampling model which establishes a correspondence between the continuous model and the case counts sampled at specific dates by looking at a cumulative number of cases between subsequent sampling dates (see \emph{Materials and Methods} and \emph{Supplementary Materials, section} \ref{app:samp_procedure}).

The results of the inference procedure are presented in Table \ref{tab:key_results} for the key model parameters $p$, $\beta_s$, and $\beta_l$ (see \emph{Supplementary Materials, section} \ref{app:synth}). In the absence of ground truth, we use the following approach to estimate the uncertainty of our inference procedure. We construct a synthetic model with a planted ground-truth parameters equal to the parameters inferred from data. Then, by generating counting noise on the same sampling dates as the ones that appear in the real data, we can construct synthetic data sets which have the same properties as the original case count data, but with the advantage that these data sets now come with a planted ground truth. We run our estimator on many instances of synthetic data with different noise realizations and compare the inference results with the planted parameters of the synthetic model. This procedure allows us to reliably estimate the uncertainty bounds of our inference procedure. Previously, a similar procedure for estimating the uncertainty in the absence of ground truth has been used in other applications involving statistical inference \cite{hannon2021real, vuffray2022programmable}. To check robustness with respect to the (unknown) counting noise, we consider the estimation error for noise generated from several families of probability distributions, and report the standard deviation results averaged over these families. The details of this procedure are given in the \emph{Supplementary Materials, section} \ref{app:synth}.

\begin{table}
\caption{\label{tab:key_results}Key model parameters inferred from the case count data, together with their single standard-deviation uncertainty averaged over several families of statistical counting noise.}
\begin{ruledtabular}
\begin{tabular}{|l|ccc|}
parameter & lower bound & inferred value & upper bound \\
\hline
$p$ & 0.041 & 0.319 & 0.597 \\
$\beta_s$ & 0.144 & 0.157 & 0.170 \\
$\beta_l$ & 0.148 & 0.155 & 0.162 \\
\end{tabular}
\end{ruledtabular}
\end{table}

\begin{figure*}
    \subfloat[Tartagal]{\includegraphics{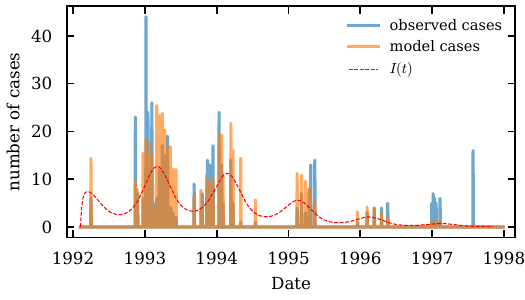}}
    \subfloat[San Ramón de la Nueva Orán (Oran)]{\includegraphics{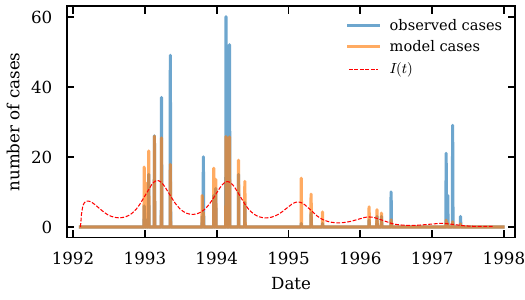}}\newline
    \subfloat[San Salvador de Jujuy (Jujuy)]{\includegraphics{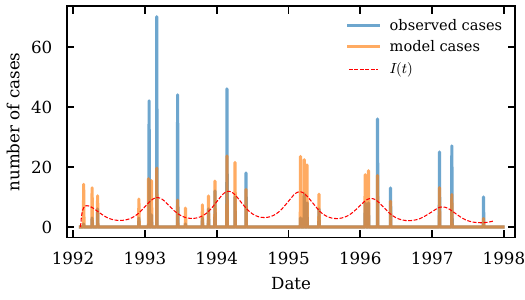}}
    \subfloat[Sampling procedure example]{\includegraphics[width=0.5\textwidth]{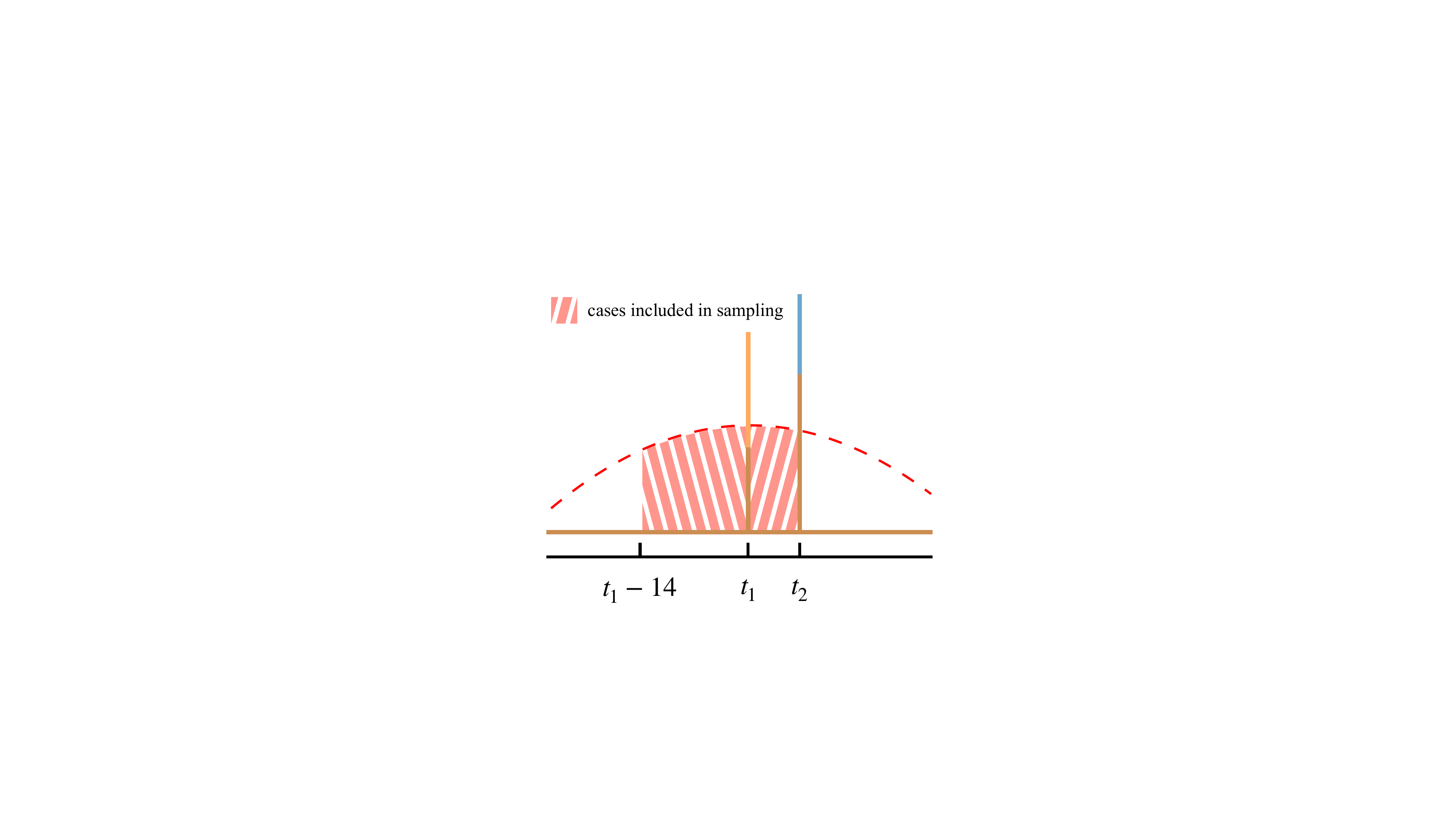}}
    \caption{\label{fig:compare} Comparison of the case count data (blue bars) and the samples obtained from the model with the inferred parameters (orange bars) for three cities: (a) Tartagal, (b) Oran, and (c) Jujuy. The red dashed line represents the number of active infected symptomatic cases according to the predictions of our continuous meta-population SIAR model. In the real data set, the case counts are not reported every day, but instead appear as discrete peaks at certain sampling days, which complicates the fitting of the continuous model. For this reason, we propose a correspondence between the continuous model and the case counts sampled at specific dates. The panel (d) explains the sampling model we use in our fitting procedure. We assume that the case counts reported on a given date correspond to a cumulative number of cases predicted by the continuous model between the minimum of the current and the previous sampling dates and 14 days before the current sampling date. This cut-off cumulative horizon represents the double of expected recovery period (fixed to 7 days, as explained in the \emph{Materials and Methods}). A more detailed explanation of the sampling procedure and the study of the impact of the choice of the sampling horizon is presented in the \emph{Supplementary Materials, sections} \ref{app:samp_procedure} and \ref{app:sampling}, respectively.}
\end{figure*}

A comparison of the true case count data and the samples obtained from the model with the inferred parameters is shown in Fig. \ref{fig:compare}.
Despite some discrepancies observed for the highest case count peaks, our simple model is able to adequately describe the variations present in the data, including the seasonal character of the cholera outbreak in Argentina.
The obtained seasonality of the transmission rate highly correlates with seasonal temperature variations in the analysed cities (see \emph{Supplementary Materials, section} \ref{app:sin}), even though such an information was not explicitly implemented in the model and was not provided to the inference algorithm.
The model suggests that cholera from this initial outbreak dies out after 1997, which is consistent with a hypothesis that the nature of the further peaks may have been significantly influenced by external factors such as Mitch hurricane in 1998 or El Niño in 1997-1998 \cite{balasubramanian2021cholera} (see \emph{Materials and Methods}).

Our results are highly robust. We successfully tested our fitting procedure with synthetic data generated with different types of noise (see \emph{Supplementary Materials, section} \ref{app:synth}).
We have also studied the scenario of a non-centered counting noise distribution corresponding to a significant under-reporting of the case counts. In the \emph{Supplementary Materials, section} \ref{app:under}, we show that this scenario is unrealistic as it leads to an unreasonably large number of predicted infected individuals, of the order of the whole city population.
Moreover, we tested sensitivity of our procedure with respect to the misspecification of the inferred migration rates (see \emph{Supplementary Materials, section} \ref{app:travel}) and different sampling horizons (see \emph{Supplementary Materials, section} \ref{app:sampling}).
Finally, we investigated how the much the length of the case count time series affected the estimated parameters (see \emph{Supplementary Materials, section} \ref{app:horizon}.
We conclude that despite many sources of potential discrepancies, estimates are robust to reasonable levels of model misspecification.

\begin{figure}
    \includegraphics[width=0.5\textwidth]{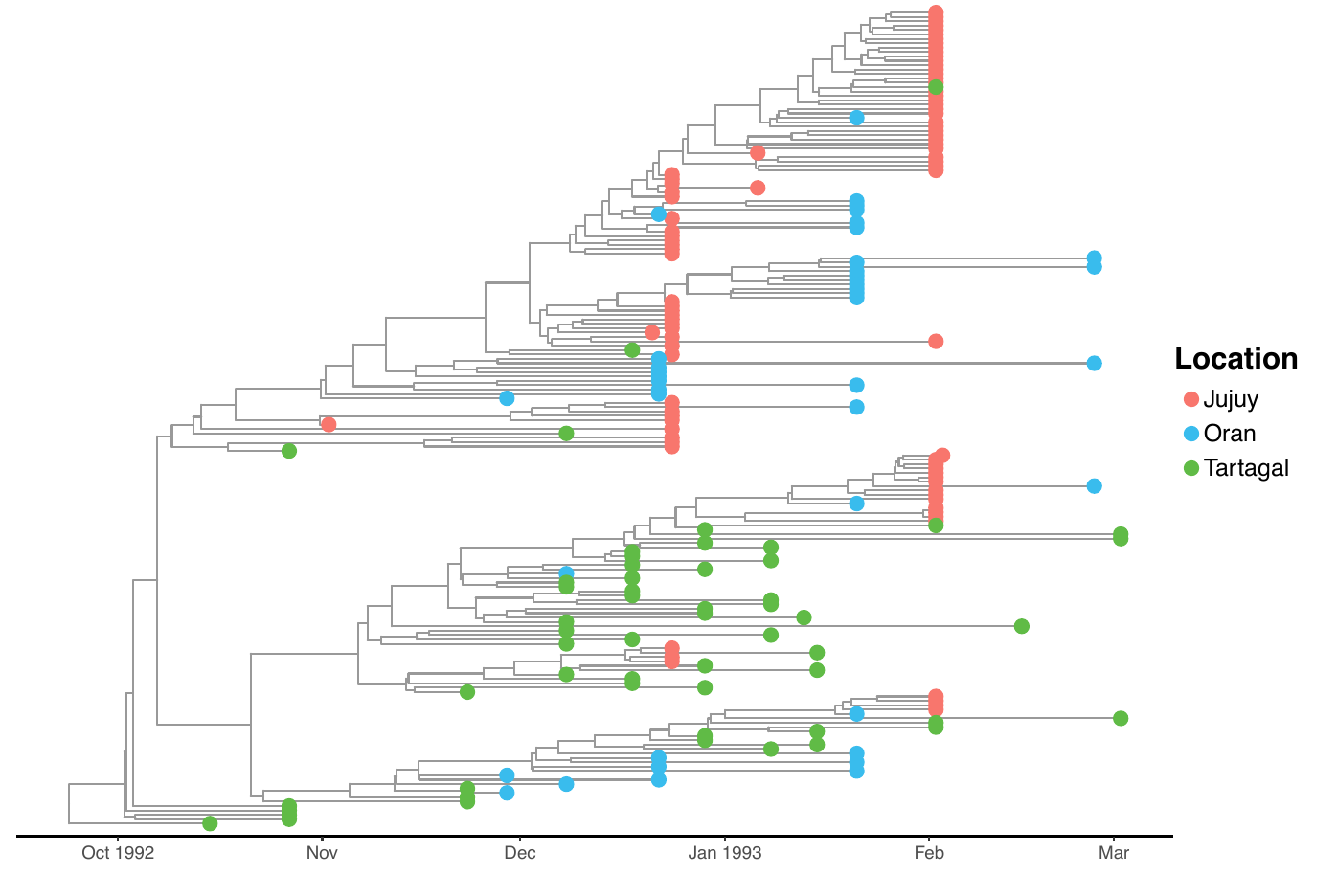}
    \caption{\label{fig:tree} The time-scaled phylogenetic tree of sequenced cholera cases sampled from Jujuy, Oran, and Tartagal. Tip colors indicate the city of sampling and position along the x-axis indicates the date of sampling. 
    The phylodynamic analysis is restricted to a single outbreak period from November 13, 1992 to March 2, 1993 that contained the majority of the sequence data (see \emph{Materials and Methods}). This maximum likelihood tree topology suggests a rapidly growing epidemic with multiple transmissions between the three locations, which suggests the key role played by the migration. A migration mechanism is included in our meta-population model through travel of asymptomatic individuals that link the dynamics in different cities.
    }
\end{figure}

{\bf Estimation of model parameters from genetic sequence data.}
We used phylodynamic analysis methods to estimate the transmission rates and the asymptomatic fraction parameter $p$ from the genetic sequence data (see \emph{Materials and Methods} and \emph{Supplementary Materials, section} \ref{app:phylogenetic_analysis} for a description of the data and the details on the inference procedure). The inferred time-scaled phylogenetic tree fixed to its maximum likelihood topology is shown in Fig. \ref{fig:tree}. The tree topology suggests an intermixed outbreak with evidence of multiple transmissions occurring between the three cities, which provides an additional justification for the inclusion of the migration mechanism in our model.
The average time between tips of the same city was 113, 128, and 145 days for Tartagal, Jujuy, and Oran, respectively, while the time between tips for different cities was 164, 160, and 145 for Tartagal-Jujuy, Tartagal-Oran, and Jujuy-Oran, respectively.
Likewise, the long branch lengths are indicative of a rapidly growing epidemic which is consistent with results from both the case-count model and the inferred transmission parameters based on the genetic sequence data, as we show next.

The principal goal of this paper consists in verifying the level of congruity of model parameters inferred from case count and genetic data. In order to check for consistency between the two data sets, we consider two different settings for inference from the genetic data. Under setting 1, we fix $p$ to the expected value inferred from the case count data, and estimate the transmission rates $\beta_s$ and $\beta_l$. Under setting 2, we leave all three parameters $\beta_s$, $\beta_l$, and $p$ free, and infer them from the genetic data. In both settings, all other parameters were set to the maximum likelihood estimates from the case-count data.
The results of the inference procedure under both settings is given in the Table \ref{tab:genetic_results_main}, along with the uncertainty bounds with 95\% highest posterior density interval (HPDI).

\begin{table}
\caption{\label{tab:genetic_results_main}Model parameters inferred from the genetic sequence data with 95\% HPDI.}
\begin{ruledtabular}
\begin{tabular}{|lc|ccc|}
setting & parameter & lower bound & inferred & upper bound \\
\hline
setting 1 & $\beta_s$ & 0.1612 & 0.1621 & 0.1629 \\
($p$ fixed)        & $\beta_l$ & 0.1596 & 0.1603 & 0.1611 \\
\hline
setting 2 & $\beta_s$ & 0.1615 & 0.1625 & 0.1634 \\
($p$ free)        & $\beta_l$ & 0.1588 & 0.1597 & 0.1605 \\
        & $p$ & 0.85 & 0.94 & 0.99 \\
\end{tabular}
\end{ruledtabular}
\end{table}

\begin{figure*}
    \subfloat[$\beta_s$ (setting 1, $p$ fixed)]{\includegraphics{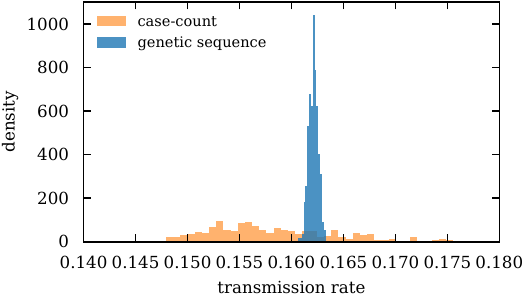}}
    \subfloat[$\beta_l$ (setting 1, $p$ fixed)]{\includegraphics{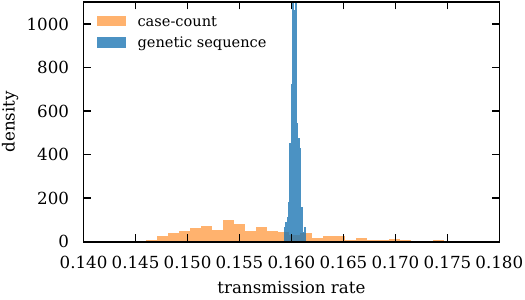}}\newline
    \subfloat[$\beta_s$ (setting 2, $p$ free)]{\includegraphics{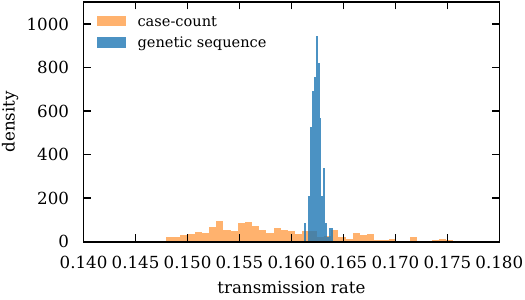}}
    \subfloat[$\beta_l$ (setting 2, $p$ free)]{\includegraphics{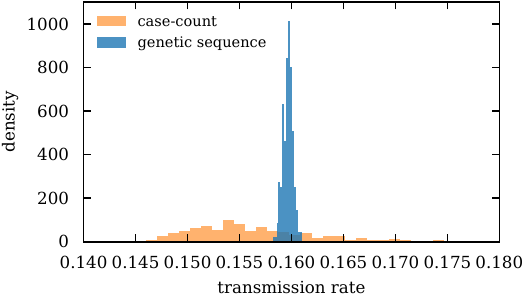}}
    \caption{\label{fig:params} Comparison of the posterior densities for $\beta_s$ and $\beta_l$ under two settings of inference from genomic data with the densities sampled from the model with parameters inferred from the case count data. 
    Under setting 1, we estimate $\beta_s$ and $\beta_l$ (the fraction of asymptomatic infections $p$ is fixed to the value inferred from the case count data), and under setting 2, we estimate $\beta_s$, $\beta_l$, as well as the free parameter $p$.
    All other parameter values were set to the maximum likelihood estimates from the case-count model.
    The orange transparent bars in the background represent the synthetic densities obtained with model parameters inferred from the case-count data. Compared to the estimates of transmission rates from the case-count model, the point estimates of $\beta_s$ and $\beta_l$ from the genetic sequence data are generally higher but still consistent within the bounds of uncertainty.}
\end{figure*}

A comparison of the posterior densities for $\beta_s$ and $\beta_l$ under both phylodynamic inference settings with the density sampled from the model with parameters inferred from the case count data is shown in Fig. \ref{fig:params}.
Compared to the estimates of $\beta_s$ and $\beta_l$ from the case-count model (see Table \ref{tab:key_results}), the point estimates from the phylodynamic model are generally higher but still consistent within the bounds of uncertainty.
We note that the uncertainty bounds from inference on the genetic data are tighter compared to the confidence intervals obtained from inference on the case count data. We believe that this is related to the fact that the inference from genetic data is much more restricted compared to the inference from case counts: all parameters besides $\beta_s$, $\beta_l$, and $p$ are fixed, and the fixed migration terms impose strong constraints on the parameter $p$. 
The phylodynamic model also infers a higher mean transmission rate $\beta_s$ for the smaller cities of Tartagal and Oran than $\beta_l$ for the larger city of Jujuy. Although the estimate of $p$ (0.94, 95\% HPDI 0.85, 0.99) under setting 2 was quite different from the case-count model, overall the $\beta_s$ and $\beta_l$ parameters were robust to fixed value of $p=0.32$ (setting 1) and the estimated value of $p=0.94$ (setting 2).
Posterior trajectories for settings 1 and 2 are discussed in the \emph{Supplementary Materials, section} \ref{app:genomic_inference_sensitivity}.

\section*{Discussion}
Our results demonstrate that phylodynamic methods applied to single-source, local cholera outbreaks produce parameter estimates that are consistent with traditional epidemiological models based on case count data.
Our study found that point estimates of the transmission rates based on case count and migration data alone produces consistently lower estimates of the transmission rates compared to the genetic sequence data.
While the estimates were well within the uncertainty bounds for estimated transmission rates from the case count data, the genetic sequence data found larger estimated outbreak sizes.
In part, this discrepancy can be due to the fact that the genetic sequence data only came from a single seasonal cycle and was therefore not constrained to maintain a sufficient susceptible population to produce outbreaks in subsequent years: we have observed a similar impact of the data length on the case-count inference in \emph{Supplementary Materials, section} \ref{app:horizon}. 
The discrepancy could also represent a real difference of opinion in the data streams.
Models based on case counts can only `see' the dynamics that are present in the observed cases, where biased or incomplete sampling can lead to underestimation of the extent of an outbreak.
However, the phylogenetic structure of pathogen genetic sequence data is fundamentally shaped by the transmission dynamics that give rise to the data regardless of how many cases were actually discovered, that is, phylodynamic methods can reveal a more complete picture of transmission dynamics given incomplete data.

The main difference in the image of the outbreak provided by these different data are the difference in uncertainty in the transmission rate estimates and the proportion of asymptomatic infections.
The high level of certainty in the genetic estimates of $\beta_s$ and $\beta_l$ should be regarded more as an artifact of necessary trade-offs made for computational efficiency, as well as of constraints on key parameters due to a smaller sample size and fixing of other model parameters.
At present, phylodynamic methods are only really feasible for relatively small data sets. 
We found that using time-variable parameters also significantly increased the computational cost, which necessitated simplifying the inference by not directly sampling phylogenetic tree topologies. 
By effectively conditioning on the maximum likelihood tree topology, we are not able to incorporate topological uncertainty in the evolutionary relationships in our sample to the estimates of the transmission rates. 
The genetic sequence data also found a much higher proportion of asymptomatic infections $p$ than the case count data.
This is partially due to limits in sampling frame for the genetic sequence data.
Because we could only fit the model to one year of sequence data, there was little cost to infecting a very large fraction of the population in a single year, which is what happened when the fraction of asymptomatic infections $p$ was not constrained (see setting 2 in Table \ref{tab:genetic_results_main}). 
We observed a similar increase in the estimate of the fraction of asymptomatic infections $p$ in the case-count inference when restricting the case-count data to one year span (see \emph{Supplementary Materials, section} \ref{app:horizon}).
The model as applied to the genetic sequence data also assumed that the migration rates were fixed to the values inferred directly from the data. 
While we found this to be a necessary assumption in our computational pipeline, it is possible that the genetic sequence data favors an overall faster migration process than is supported by the travel data and compensates by simply assuming a higher asymptomatic fraction: we observed a similar effect in our analysis of sensitivity to a misspecification of migration flows $f_{ij}$ reported in \emph{Supplementary Materials, section} \ref{app:travel}.
Regardless of a particular inferred or assumed value of $p$, we found that the transmission amplitudes $\beta_s$ and $\beta_p$ were very robust to different values of the fraction of asymptomatic infections (see Table \ref{tab:genetic_results_main}).

In this paper, we demonstrated that, taken independently, case count and genetic sequence data provide a complementary view of a cholera transmission dynamics at the city/regional-level.
This analysis represents a necessary data consistency check prior to building a joint framework for inference from different data sources.
The next challenge for integrating genetic sequence data into cholera modeling projects is joint inference on both case count time series and genetic sequence data.
Allowing both case count and genetic sequence data to contribute to a model would allow us to better estimate parameters such as the level of under-reporting of cholera cases (referred to as statistical counting noise in \emph{Supplementary Materials, sections} \ref{app:synth} and \ref{app:under} in our study), the fraction of asymptomatic infections $p$, and possible differences in asymptomatic versus symptomatic transmission at the population level.
Developing a population of generalized cholera models that can integrate case count, genetic sequence, and migration data into a single picture of cholera outbreak dynamics will give cholera researchers more power to differentiate between different modeling assumptions, for example, more carefully elucidating the role of environmental versus direct transmission in sustaining cholera transmission in multi-year outbreaks.

\section*{Materials and Methods}
{\bf Fixed model parameters.}
We infer the free parameters of the model from the data. However, some of the parameters are not independent, or are well documented, and hence can be fixed. The model parameters that are fixed during our inference procedure include the recovery rate $\gamma$, the city populations $N_i$, the initial values of compartments $I_i(0)$, $R_i(0)$ and the migration parameters $f_{ij}$.
The recovery rate of cholera is well documented in the literature \cite{hendrix1971pathophysiology, weil2009clinical, weil2014bacterial, mondiale2017weekly} and equal to $\frac{1}{7}$, which represents the average period of $7$ days to recovery.
We use available historical population data to set the total population size $N$ of each city: $N = 4.4 \cdot 10^4$ in Tartagal, $N = 5.1 \cdot 10^4$ in San Ramón de la Nueva Orán and $N = 2 \cdot 10^5$ in San Salvador de Jujuy. 
The initial values of $I_i(0)$ and $R_i(0)$ for all cities $i$ are taken as zero, since we follow the assumption that the infection was most likely introduced in Salta and Jujuy provinces by a single asymptomatic migration event \cite{mazzafero1995epidemic}.
As a consequence, for all locations $i$ in the meta-population model, $S_i(0) = N_i - A_i(0)$, where $A_i(0)$ is a free parameter which is inferred from the case count data, but is assumed to be the same for all three cities, i.e. $A_i(0) = A(0) \, \forall i$.
Further following the documented suggested sequence of events preceding the initial epidemic outburst in Salta and Jujuy provinces \cite{mazzafero1995epidemic}, we assume that $5 \leq A(0) \leq 15$.
Lastly, the migration rates $f_{ij}$ for a pair of locations $(i,j)$ are fixed based on available documented and extrapolated data about domestic travel in Argentina.
Values of $f_{ij}$ used in the model can be found in the \emph{Supplementary Materials, section} \ref{app:fij}.

In addition, a certain choice for the starting and ending dates is needed to connect the model predictions with the available case count data.
The first case in Salta province was sampled on February 10th 1992 in our data set, whereas according to \cite{mazzafero1995epidemic}, the migration of asymptomatic cases into Salta province started around the first week of February 1992.
As a result, we use an intermediate date of February 8th 1992 as the starting date in our inference procedure.
The last cases sampled in Argentina and present in our data set are from 2002 (2000 in Salta province), however, as suggested in \cite{balasubramanian2021cholera}, major external events such as Mitch hurricane in 1998 or El Niño in 1997-1998 significantly affected the dynamics of cholera in South America.
In order to omit this potentially strong influence of the external events on our data, we only use the available data up until 1998.

{\bf Inference of model parameters from case count data.}
Epidemiological case data is subject to different sources of uncertainty and statistical noise: missing observations, delayed reporting, technical and laboratory errors, etc.
While the observed number of cases can be higher or lower than the true number of infected individuals in a given period, it is reasonable to assume that there exists a delay between the occurrence and observation of a case for both objective (e.g., lack of nearby facilities) and administrative (e.g., delay in reporting) reasons. As a consequence of such a reporting delay, the case counts appearing in the data appear as discrete peaks at specific days. To account for this delay, we assume that in each city, raw data represents a sampling of its infected population, where the number of cases reported on a specific date consists of a cumulative number of symptomatic cases observed between the current and the previous sampling dates, unless the previous sampling date appears more than 14 days before the current data (in which case the sampling horizon is fixed to 14 days). This cut-off cumulative horizon represents the double of expected recovery period $1/\gamma$ (which is fixed to 7 days, as explained above). A graphical representation of this procedure is shown in Fig. \ref{fig:compare}(d).
Details of the assumed sampling procedure are described in \emph{Supplementary Materials, section} \ref{app:samp_procedure} and additional experiments showing the impact of choosing different sampling horizons is presented in \emph{Supplementary Materials, section} \ref{app:sampling}.
This procedure allows us to directly compare the reported case data with the predictions of our continuous model using the sampling procedure described above.

In most of presented results, we infer the model parameters by minimising the average square error between the modeled and observed case counts.
In a maximum likelihood setting, this is equivalent to the assumption of normally-distributed statistical noise on the case counts. In our analysis with synthetic data, we found this assumption to be well justified (\emph{Supplementary Materials, section} \ref{app:synth}).
In our inference procedure, the average square error is minimized using the Limited-memory Broyden–Fletcher–Goldfarb–Shanno algorithm with bounded constraints (L-BFGS-B) \cite{byrd1995limited}.
Given that the resulting optimization problem is highly non-convex, we used the warm-starting strategy by initializing the algorithm multiple times, with different initial conditions, in order to increase the probability of finding the global minimum.
The code used to obtain all our results is available at \cite{inference-code-data}.

{\bf Inference of model parameters from pathogen sequence data.}
Publicly available alignment files from a previously published study on whole genome sequencing of \textit{Vibrio cholerae} isolates from Argentina \cite{dorman2020genomics} were downloaded from FigShare \cite{genetic-data}.
Here, we utilized the alignment data using the Peruvian strain A1552 isolated in 1991 as the reference genome \cite{dorman2020genomics}.
We used the PhyDyn library \cite{volz2018bayesian} for the BEAST2 phylogentics platform \cite{bouckaert2019beast} to fit the cholera ODE model to the genetic sequence data, see \emph{Supplementary Materials, section} \ref{app:phylogenetic_analysis} for implementation details.  
We restrict the phylodynamic analysis to a single outbreak period from November 13, 1992 to March 2, 1993 that contained the majority of the sequence data, resulting in 55, 41, and 90 sequences for Tartagal, San Ramón de la Nueva Orán, and San Salvador de Jujuy respectively.
Although we set the starting time for the outbreak $t_0$ to February 8th, 1992 to be consistent with the case count model, we exclude the 11 sequences sampled between in March and April of 1992 (see \emph{Supplementary Materials, section} \ref{app:samples}) to improve computational efficiency.  
We use the known sampling times of each sequence with the HKY substitution model \cite{hasegawa1985dating} assuming a strict clock to time scale the tree.
To simplify the inference, which we found to be inefficient when sampling both tree and model space jointly, we fix the tree topology to its maximum likelihood topology inferred using IQtree2 \cite{minh2020iq}.  

In order to check a consistency between the phylogenetic and traditional epidemiological data, we use two different settings for inference from the genomic data. Under the setting 1, we fix $p$ to the expected value inferred from the case count data, and estimate the transmission rates $\beta_s$ and $\beta_l$. Under the setting 2, we infer all three parameters $\beta_s$, $\beta_l$, and $p$ from the genomic data. Under both settings, inference procedures were run in such a way that all effective sample sizes were greater than 700 with the first 10\% of samples being removed, and all remaining parameters were set to the maximum likelihood estimates from the case-count data. The details on phylodynamic inference under both settings are provided in the \emph{Supplementary Materials, section} \ref{app:genomic_inference_sensitivity}.

\section*{Acknowledgments}
Authors acknowledge support from the Laboratory Directed Research and Development program of Los Alamos National Laboratory under projects numbers 20200121ER (MW, LC, JK, CM, ERS, AYL) and 20210529CR (MW), as well as from the program NIH NCATS KL2TR001448 (DD). We thank Matthew Dorman for his comments on the manuscript.\\

\noindent{\bf Data availability}
The case count data used in this study is available at \cite{inference-code-data}. Publicly available alignment files from a previously published study on whole genome sequencing of \textit{Vibrio cholerae} isolates from Argentina \cite{dorman2020genomics} were downloaded from FigShare \cite{genetic-data}. In this study, we utilized the alignment data using the Peruvian strain A1552 isolated in 1991 as the reference genome. All other data that support the plots within this paper and other findings of this study are available from the authors on reasonable request.\\

\noindent{\bf Code availability}
The code implementing inference from case count data is available at \cite{inference-code-data}.\\

\noindent{\bf Author contributions} MW, LC, CM, ERS, DD, and AYL designed the research. JC and DD provided high-fidelity case count and genomic sequence data. MW and JK collected the historical migration data and performed an extrapolation. MW, LC, ERS, and AYL developed the mathematical methods. MW performed the numerical analysis using case count data. LC performed the numerical analysis using the genomic sequence data. MW, LC, ERS, DD, and AYL wrote the manuscript. All authors proofread and commented on the manuscript.\\

\noindent{\bf Competing interests} The authors declare no competing interests.\\

\noindent{\bf Additional information}  
This work was reviewed by the UNM Human Subjects Research Program under Study ID 19-484 and by the LANL Human Subjects Research Review Board under study ID LANL 19-16 E.\\

\noindent{\bf Supplementary Information} is available for this paper.\\

\noindent{\bf Correspondence and requests for materials} should be addressed to DD and AYL.

\bibliographystyle{naturemag}
\bibliography{literature}
\balancecolsandclearpage

\onecolumngrid
\begin{center}
{\large Supplementary Materials}
\end{center}

\twocolumngrid

\appendix
\renewcommand{\thefigure}{S\arabic{figure}}

\setcounter{figure}{0}

\renewcommand{\thetable}{S\arabic{table}}

\setcounter{table}{0}

\section{Sample sizes for case count data and pathogen sequence data}
\label{app:samples}
\begin{figure}[!h]
    \subfloat[Tartagal]{\includegraphics{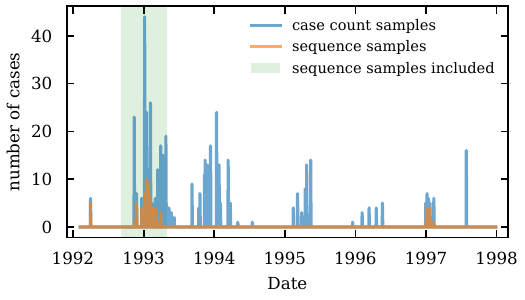}}\newline
    \subfloat[San Ramón de la Nueva Orán (Oran)]{\includegraphics{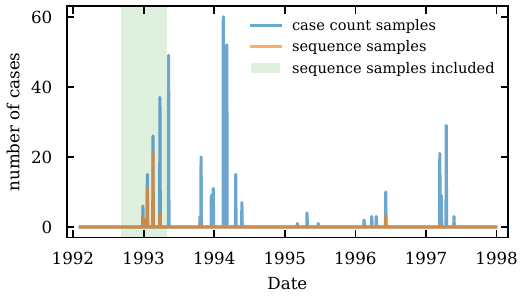}}\newline
    \subfloat[San Salvador de Jujuy (Jujuy)]{\includegraphics{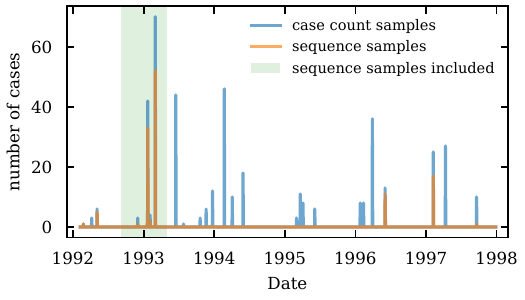}}
    \caption{\label{fig:samples} Number of all cases taken into account in the case count analysis, along with the number of cases with associated sequenced genetic data (samples from both sources are overlaid). Light green background highlights the period used in the phylogenetic analysis.}
\end{figure}

As described in the main text, only samples with certain descriptive data were taken into account in the case count analysis.
Not all of them were connected with the sequenced genetic data.
Since our goal is to compare two distinct approaches, where one is based only on case count data, while the other on phylogenetic analysis, we narrow the data-set to only three cities, where the number of samples is sufficient for both methods.
A summary of all available cases in the final curated data-set is shown in Fig. \ref{fig:samples}.
For the sequenced genetic data, we further limit the sample to those isolated between 8th September 1992 and 30th April 1993.
Compared to other years, the samples collected during this seasonal outbreak had the most genetic diversity across the three locations, which is required to fit a meta-population phylodynamic model.

\section{Computation of migration flow parameters $f_{ij}$}
\label{app:fij}
The two principle means of transportation in Argentina are buses and flights.
Trains represent another significant mean of transportation, however, they are not used on routes connecting the cities of our interest.
Historical data relevant to the dates of interest to our study is not directly available, and hence we use the following extrapolation procedure to estimate the migration rates using available historical data in other time periods.
According to \cite{aereo2017}, the number of people travelling in 2017 by plane from Salta province to Jujuy province was equal to around 15,000 passengers.
As for buses, \cite{bus2017} reports 520,000 passengers travelling between Salta and Jujuy provinces in 2017, and 330,000 passengers travelling inside Salta province in the same year.
For our purposes, we need to estimate similar passenger flows during the period relevant to our study. In order to re-scale these numbers, so that they would better reflect values between 1992 and 1998, we use data from \cite{bus-rescale} and \cite{aereo-rescale} to compute the ratio between the the overall number of passengers travelling in 1993 and 2017, separately for planes and buses.
Subsequently, we multiply the above numbers by these ratios, which are equal to $1.0$ and $0.4$ for planes and buses accordingly.
By summing the above re-scaled values for buses and planes, we obtain yearly traveling matrix for provinces.
Finally, we divide it by $365$ to get daily rates denoted as $g_{ij}$ (where $i$ and $j$ refer to provinces).

Since the data is aggregated to provinces, we re-scale it to city level by normalising each direction by the cities populations divided by the provinces populations. More formally, we use the following relations for the migration flow parameters:
\begin{equation}
    f_{ij} = \frac{n_i n_j}{m_{k_i} m_{k_j}} \cdot g_{k_i k_j},
\end{equation}
where $g_{k_i k_j}$ is the number of people travelling daily from province $k_i$ (province of city $i$) to province $k_j$ (province of city $j$); $n_i$ is the population of city $i$; and $m_{k_i}$ is the population of province $k_i$ (province of city $i$).
Using the historical numbers from \cite{city-population} and \cite{province-population}, we compute the ratios $n_i n_j / m_{k_i} m_{k_j}$, which are reported in Table \ref{tab:scale}.

\begin{table}
\caption{\label{tab:scale}Population re-scaling ratios $n_i n_j / m_{k_i} m_{k_j}$ for migration flows between three analysed cities.}
\begin{ruledtabular}
\begin{tabular}{|l|ccc|}
City & Tartagal & Oran & Jujuy \\
\hline
Tartagal & 0.0000 & 0.0030 & 0.0180 \\
Oran & 0.0030 & 0.0000 & 0.0216 \\
Jujuy & 0.0180 & 0.0216 & 0.0000 \\
\end{tabular}
\end{ruledtabular}
\end{table}

Joining Table \ref{tab:scale} and matrix $g_{k_i k_j}$ leads to the values presented in Table \ref{tab:fij}, which are subsequently used in our meta-population model, as explained in section \ref{app:model}. Given that this extrapolation may produce biased values of migration flows, in section \ref{app:travel} below, we study the robustness of inference using our meta-population model with respect to the misspecification of migration rates.

\begin{table}
\caption{\label{tab:fij} $f_{ij}$ values, representing the average number of people traveling daily between the analysed cities.}
\begin{ruledtabular}
\begin{tabular}{|l|ccc|}
City & Tartagal & Oran & Jujuy \\
\hline
Tartagal & 0.00 & 2.74 & 26.50 \\
Oran & 2.74 & 0.00 & 30.71 \\
Jujuy & 26.50 & 30.71 & 0.00 \\
\end{tabular}
\end{ruledtabular}
\end{table}

\section{Details of the ODE cholera model}
\label{app:model}
We propose a Susceptible-Infected-Asymptomatic-Recovered (SIAR) meta-population model of cholera transmission dynamics, where each city is modelled through a set of ordinary differential equations.
The cities are coupled through migration terms, which allow the susceptible and asymptomatic individuals to travel between the cities, but properly normalized in such a way that the populations of the cities remain constant.
More precisely, we use the following set of equations:
\begin{widetext}
\begin{equation}
\label{eq:ode}
\begin{split}
    \frac{dS_i}{dt} &= -\beta_i(t) \cdot (I_i + A_i) \cdot S_i - \sum_{j \neq i} f_{ij} \cdot \left( \frac{A_j}{S_j + A_j + R_j} - \frac{A_i}{S_i + A_i + R_i} \right),\\
    \frac{dI_i}{dt} &= (1 - p) \cdot \beta_i(t) \cdot (I_i + A_i) \cdot S_i - \gamma I_i,\\
    \frac{dA_i}{dt} &= p \cdot \beta_i(t) \cdot (I_i + A_i) \cdot S_i - \gamma A_i + \sum_{j \neq i} f_{ij} \cdot \left( \frac{A_j}{S_j + A_j + R_j} - \frac{A_i}{S_i + A_i + R_i} \right),\\
    \frac{dR_i}{dt} &= \gamma \cdot (I_i + A_i).
\end{split}
\end{equation}
\end{widetext}
The set of equations is the same for each city (indexed by $i$), but some of the parameters (indexed by $i$) may differ from location to location.
The initial conditions are given by $S_i(0)$, $I_i(0)$, $A_i(0)$ and $R_i(0)$.
At each time point $t$ the variables are related by $S_i(t) + I_i(t) + A_i(t) + R_i(t) = N_i$, where $N_i$ is the population of city $i$.
Below, we discuss the meaning behind model parameters.

\,

\noindent \textbf{$\gamma$ parameter}

\noindent The recovery rate for cholera is well documented \cite{hendrix1971pathophysiology, weil2009clinical, weil2014bacterial, mondiale2017weekly}, and for this reason we keep it fixed as $\frac{1}{7}$, which represents the expected 7 days to recovery. It is set to the same value for each city.

\,

\noindent \textbf{$p$ parameter}

\noindent The fraction of infected cases falling into the asymptomatic compartment upon the infection process.
The value is the same for each city and it is within the $(0, 1)$ interval.

\,

\noindent \textbf{$\beta_i(t)$ function}

\noindent For each location $i$, the transmission parameter is described by the following equation:
\begin{equation}
    \beta_i(t) = \beta_i \cdot \frac{\sin(2 \pi \frac{t}{365} + \pi + \phi) + 1 + m_0}{2 + m_0},
    \label{eq:beta}
\end{equation}
where $\beta_i$ represents the transmission amplitude, and the rest of the expression models an effecitive influence of the seasonality. In particular, $\phi$ is the phase of the seasonality (relative to $2 \pi$ representing the whole year), and $m_0$ is an offset that fixes the minimum value of the transmission parameter.
All three parameters are positive and $\phi$ is also smaller or equal to $2 \pi$.
We assume that $\beta_i$ can be different for smaller (Tartagal and Oran) and larger (Jujuy) cities in our study.
For instance, some discrepancy could be attributed to variations in city infrastructure.
On the other hand, we assume that parameters $\phi$ and $m_0$ are common to all three cities.

Note that with this parametrization, we assume a yearly seasonality $\beta_i(t)$, which is in line with \cite{daisy2020developing} and what can be visually observed in the case count data in Fig. \ref{fig:samples}. Although the precise mechanism of seasonality influence on cholera dynamics remains an open question, it is generally agreed that cholera transmission dynamics correlates with seasonal conditions \cite{emch2008seasonality}.
A visual inspection of the case counts time variation in Fig. \ref{fig:samples} also shows a significant increase in the number of cases during the warmer periods of the year, while displaying a relatively small number of cases during the cooler months.
In our modeling approach, we do not make any explicit causal hypothesis on the effect of seasonality on cholera dynamics. Instead, similarly to our modeling choice where we do not explicitly include an environmental compartment with associated additional parameters, we choose to absorb the seasonality as a periodic functional dependence on the \emph{effective} transmission parameters $\beta_i(t)$. Previously, a similar inclusion of seasonal dependence inside the effective transmission parameters has been considered in \cite{emvudu2013mathematical}.
A particular choice for the periodic function to describe the seasonality is not important, and we have chosen a sinusoidal dependence for its parametric simplicity.
In prior work \cite{kolaye2019mathematical}, seasonality has been included in the model in a different way, unrelated to direct transmission, but through the environmental factors (bacterial reservoirs capacity), and was directly dependent on the average temperature data.
We do not directly include the historical temperature data in our model, and instead fix the seasonality related parameters directly from data.
During the fitting procedure, we only fix the period of the functional dependence (matching it to a year), while leaving the amplitude, the average value and most importantly, the phase, as free parameters that are reconstructed as a part of the fit from case counts. In what follows, we compare the inferred value of this seasonal element with historic yearly temperature variations in section \ref{app:sin}.

\,

\noindent \textbf{$f_{ij}$ parameters}

\noindent Finally, the migration-related parameters $f_{ij}$ represent the number of people travelling daily from city $i$ to city $j$.
We assume that infected (symptomatic) people do not travel and that the migration matrix is symmetric.
Although the symmetry of the migration matrix is not required, this property has the benefit of keeping the city populations constant over time.
For the same reason, since the fraction of susceptible or individuals individuals among all travelers is \emph{a priori} unknown, the change in the number of asymptomatic needs to be balanced with a change in the number of susceptibles (see Eq. (\ref{eq:ode})).
We fix the values of the migration flow parameters based on the available and extrapolated historical travel data, as explained in detail in section \ref{app:fij}.

\section{Sampling procedure for case-count predictions}
\label{app:samp_procedure}
We formally describe the procedure for sampling discrete case counts from the continuous model, described in the main text.
Let us denote the $k$th sampling date for city $i$ as $t_k^i$ and the number of cases sampled on that day as $\bar{G}(t_k^i)$.
Then we propose the following model for case counts on a specific date which includes the cumulative number of cases prior to that date, as well as the statistical counting noise:
\begin{equation}
    \bar{G}(t_k^i) = \int_{\hat{t}_k^i}^{t_k^i} dG_i + \eta_k^i,
    \label{eq:sample}
\end{equation}
where $\hat{t}_k^i = \max( t_{k-1}^i, t_k^i - 14 )$ and
\begin{equation}
    dG_i = (1 - p) \cdot \beta_i(t) \cdot (I_i + A_i) \cdot S_i dt.
\end{equation}
$G_i(t)$ has the meaning of the number of all symptomatic infected cases that appeared up until time $t$, in city $i$.
$\bar{G}(t_k^i)$ represents the final sampled value, the integral $\int_{\hat{t}_k^i}^{t_k^i} dG_i$ is the prediction of the model, and $\eta_k^i$ is the statistical noise that reflects many independent factors such as missing observations, delayed reporting, technical and laboratory errors, \emph{etc}.

\section{Inference with real and synthetic case-count data}
\label{app:synth}
Following on the definitions introduced in previous section \ref{app:samp_procedure}, let us for simplicity denote the observed data sample at a specific date $t_k^i$ as $x_k^i = \bar{G}(t_k^i)$, and the sample predicted by the model as $y_k^i = \int_{t_k^i-14}^{t_k^i} dG_i$.
As described in the main text and in \emph{Materials and Methods}, our inference procedure is based on minimising the square difference between the observed samples and the ones predicted by the model.
More formally, we minimize the following expression:
\begin{equation}
\label{eq:likelihood}
    l\left(\beta_s, \beta_l, m_0, p, \phi, A(0) | x\right) = \sum_i \sum_k \left(x_k^i - y_k^i\right)^2.
\end{equation}
In the case of independent and identically distributed Gaussian noise $\eta_k^i$, this least-squares estimator corresponds to the negative log-likelihood that the model with parameters $\{\beta_s, \beta_l, m_0, p, \phi, A(0)\}$ produces samples $\{y_k^i\}_{i,k}$. Parameters obtained by minimising the expression \eqref{eq:likelihood} using the available case-count data are given in Table \ref{tab:all_results}.

\begin{table}
\caption{\label{tab:all_results} Model parameters inferred from the case-count data, along with their single standard-deviation uncertainty.}
\begin{ruledtabular}
\begin{tabular}{|l|ccc|}
parameter & lower bound & inferred value & upper bound \\
\hline
$p$ & 0.041 & 0.319 & 0.597 \\
$\beta_s$ & 0.144 & 0.157 & 0.170 \\
$\beta_l$ & 0.148 & 0.155 & 0.162 \\
$\phi$ & 4.02 & 5.81 & 7.60 \\
$m_0$ & 3.2 & 11.0 & 18.8 \\
$A(0)$ & 6.4 & 10.0 & 13.6 \\
\end{tabular}
\end{ruledtabular}
\end{table}

As discussed in the main text, in reality, the true distribution of the statistical counting noise $\eta_k^i$ in Eq. (\ref{eq:sample}) is unknown, given a small sample size and a large number of factors affecting the reporting of case counts. However, we can make several reasonable assumptions on the noise distribution. First, the observed sample $\bar{G}(t_k^i)$ cannot be negative, which imposes restriction on the value of the noise realization $\eta_k^i$. Second, we assume that the statistical noise distribution is centered around zero, which accounts for both under-counting and over-counting (in section \ref{app:under}, we consider a popular scenario of under-counting corresponding to a non-centered counting noise, and show that this leads to non-reasonable values of the inferred parameters). Third, we expect that the variance of the noise distribution adequately describes fluctuations observed in the case count data.

In order to test the robustness of our approach to the noise distribution, we run a series of synthetic simulations, where we use several families of noise distributions (Gaussian, negative binomial and gamma distributions) that satisfy these assumptions.
For each sampling date we generate a number of candidate observed cases using a proposed noise distribution.
The parameters of the distribution is fixed in such a way that the expected value of the predicted cases is equal to the model prediction in the absence of the noise (which is known for synthetic simulations), while the variance is the same as for the original data (around 182).
Unless specified otherwise, the model parameters are set to the values inferred from the case count data.

\begin{figure}[b]
    \includegraphics{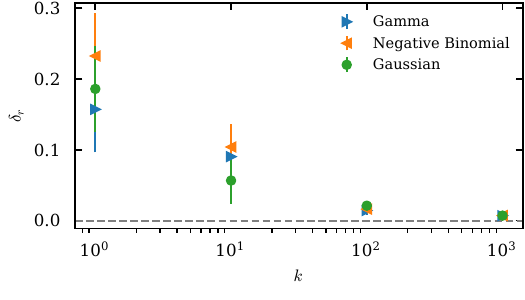}
    \caption{\label{fig:conv} The average relative error $\delta_r$ on inferred parameters as a function of the number of replicas $k$ of synthetic data for different noise distributions -- blue triangles for gamma distribution, yellow triangles for negative binomial distribution and green circles for Gaussian distribution.
    The inference was repeated for 8 different realisations of the noise, allowing to draw standard deviations for each $k$ and each noise distribution.}
\end{figure}

The goal of the first experiment is to test inference quality in a synthetic setting with a planted ground truth. This experiment will inform us on the quality of the inferred solution, as well as on the expected accuracy under the available amount of data under different possible distributions of counting noise.
Ideally, we prefer to preserve the sampling dates present in the real data set. However, to test data requirements, we would like to be able to increase the number of samples, so that we can understand the method's rate of convergence to the true solution.
In order to achieve both goals, we produce $k$ replicas of synthetic data with different realisations of the described synthetic noise.
This way, we obtain $k \cdot n$ data points, where $n$ is the sample size in the original data (number of samples in all three analysed cities) and $k$ is the number of replicated samples.
Then, we fit the model to multiple sampling realisations at the same time.
By controlling $k$ we test whether the inferred parameters converge to the true planted parameter values as the effective sample size grows.
We define the relative $\ell_1$ error $\delta_r$ on the inferred parameters as follows:
\begin{widetext}
\begin{equation}
    \delta_r = \frac{|\beta^{*}_s - \beta_s| + |\beta^{*}_l - \beta_l| + |m_0^* - m_0| + |p^* - p| + |\phi^* - \phi| + |A(0)^* - A(0)|}{\beta^{*}_s + \beta^{*}_l + m_0^* + p^* + \phi^* + A(0)^*},
\end{equation}
\end{widetext}
where parameters marked with an asterisk are the planted ground-truth ones used to generate synthetic data, and the parameters without an asterisk are the inferred ones.

Fig. \ref{fig:conv} shows the average (over 8 sets of noise realizations) relative error $\delta_r$ on inferred parameters as a function of $k$. For comparison, we use Gaussian counting noise (in which case the estimator in \eqref{eq:likelihood} corresponds to a maximum likelihood approach), as well as binomial and gamma noise distributions. We see that the quality of the inferred solution is very robust even in the case of the latter distributions which are strongly non-Gaussian (non-symmetric, positive-support densities).

The second experiment is designed with the goal of estimating the uncertainty of the obtained results in the synthetic setting. To this end, we first generate synthetic data using the parameters inferred from original data (uaing both gamma and negative binomial distributions), and then apply our inference approach to recover the mentioned parameters.
We repeat this procedure multiple times and compute the variance of the obtained estimates.
Note that similarly to the inference from the real data, no uncertainty is associated with the recovery parameter $\gamma$, since this parameter is fixed in our inference procedure.
The uncertainty obtained with the above procedure and averaged over different distributions is reported in Table \ref{tab:all_results} and main text Table \ref{tab:key_results}.

\section{Seasonality effects in the inferred effective transmission parameters $\beta_i(t)$}
\label{app:sin}
\begin{figure}
    \subfloat[Salta province]{\includegraphics{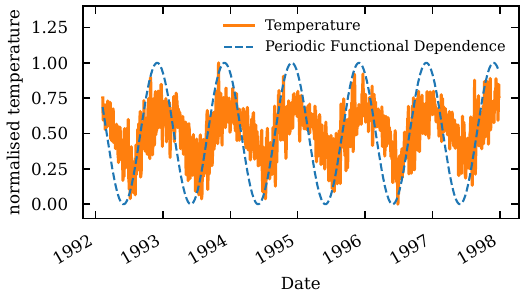}}\newline
    \subfloat[Jujuy province]{\includegraphics{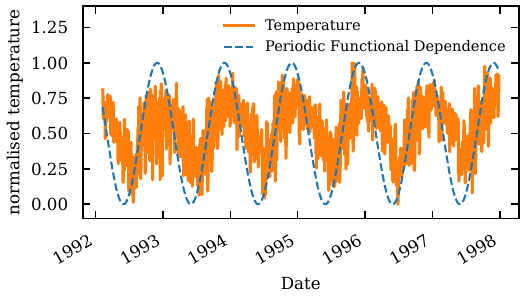}}
    \caption{\label{fig:temp} Historical data on average daily temperature (orange solid line) in Salta and Jujuy provinces, along with $\beta_i(t)$ sinusoidal functional dependence controlled by $\phi$ and $m_0$ (dashed blue line) inferred from the case count data. For comparison purposes, both quantities are renormalised to fit inside of the $(0, 1)$ interval.}
\end{figure}

In this section, we compare the inferred periodic component of $\beta_i(t)$ controlled by the parameters $\phi$ and $m_0$ with historical data on temperature variations.
In Fig. \ref{fig:temp}, we show that the resulting inferred phase (see Table \ref{tab:all_results}) correlates well to the variations of temperature, especially in the periods of rising temperature, although our model did not take temperature as an explicit input parameter.

\section{Examination of the scenario of under-reporting of the case counts}
\label{app:under}

It is often claimed that due to various objective reasons, available case-count data is often under-reported. In this section, we study a case of a significant under-reporting scenario corresponding to the case of non-centered counting noise. We show that this assumption would lead to unreasonable values of model parameters.

One possibility to include the assumption of purely negative counting noise -- which directly translates into strict under-counting in the sampling procedure -- consists in modifying the objective function in Eq. (\ref{eq:likelihood}).
An appropriate modification would forbid the events $x_k^i > y_k^i$ for any $k$ and $i$, for instance by setting the likelihood to $+\infty$ every time the model parameters produce at least one such inequality.
It is easy to see that such a strong condition makes it impossible to infer the parameters that would adequately describe that data: the resulting trajectory cover values already in the first peak of epidemic in 1993 (see Fig. \ref{fig:samples}) would result in an infection of the whole city population, which is not consistent with leaving an sufficient susceptibility pool to sustain a multi-year outbreak.

\begin{figure*}
    \centering
    \subfloat[Tartagal]{\includegraphics{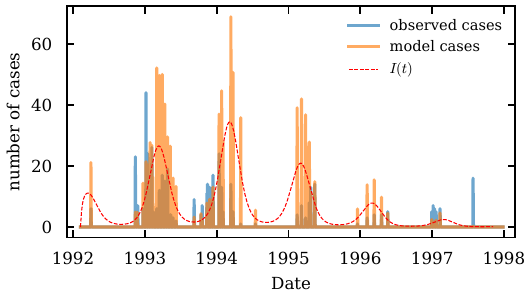}}\\
    \subfloat[San Ramón de la Nueva Orán (Oran)]{\includegraphics{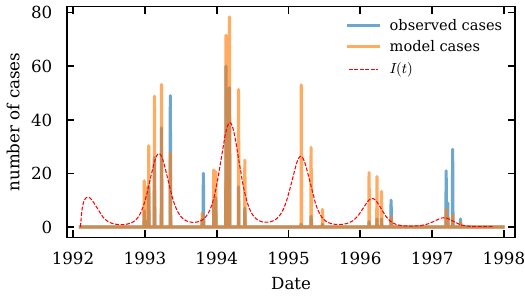}}\\
    \subfloat[San Salvador de Jujuy (Jujuy)]{\includegraphics{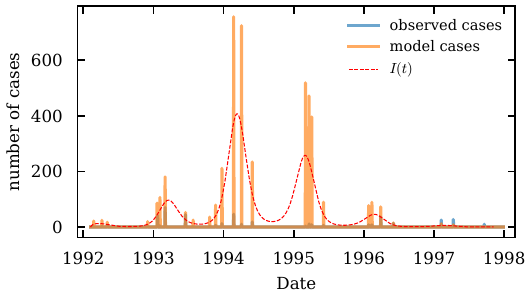}}\\
    \caption{\label{fig:undercounting} A comparison of the original case count data (blue lines) and the samples obtained from the model with the parameters inferred using Eq. (\ref{eq:under_likelihood}) with $M = 20$, strongly favoring under-counting of the case counts (orange lines).
    The red dashed line represents the number of active infected (symptomatic) people, according to the continuous model with the same parameters. A relatively poor fit suggests that according to our dynamic model, the case count data does not seem to be strongly under-reported.}
\end{figure*}

A softer under-counting condition may be achieved by a simple modification of Eq. (\ref{eq:likelihood}) which favors under-counting:
\begin{equation}\label{eq:under_likelihood}
    \begin{split}
        l(\{\beta_i^0\},& m_0, p, \phi, A(0) | x) = \\
        &\begin{cases}
            \sum_i \sum_k |x_k^i - y_k^i|^M, & \text{if } x_k^i > y_k^i \\
            \sum_i \sum_k (x_k^i - y_k^i)^2, & \text{otherwise,}
        \end{cases},
    \end{split}
\end{equation}
where $M$ is a parameter which controls the preference for negative counting noise values.
Note that $M \rightarrow +\infty$ is equivalent to the previously discussed scenario of strict under-counting, here interpreted as an infinite barrier.
A comparison between case count data and a model obtained according to the objective in Eq. (\ref{eq:under_likelihood}) with $M = 20$, is shown in Fig. \ref{fig:undercounting}.
Although for finite $M$ it is possible to reproduce seasonal peaks appearing in the data (see Fig. \ref{fig:samples}), we observe a poor fit of the case count data.
This is specifically apparent for San Salvador de Jujuy, where the model suggests number of cases reaching 700 cases for some of the sampling dates, while in the original data the case counts do not exceed 70 at any point of time.
Additionally, a model inferred using Eq. \eqref{eq:under_likelihood} with $M = 20$ predicts that 60\% of the city population was symptomatically  infected during the outbreak.
Apart from these inconsistencies, the inference procedure consistently ends up with a prediction $p = 0$ for values of $M$ greater than 2 when the objective function \eqref{eq:under_likelihood} is used, which is clearly unrealistic since it corresponds to the absence of asymptomatic individuals. This justifies the assumption on centered counting noise accounting for both under-counting and over-counting that we used in section \ref{app:synth}.

\section{Sensitivity of the inference procedure to the misspecification of migration rates}
\label{app:travel}
In section \ref{app:fij} we provided a detailed description on the estimation of the migration rates $f_{ij}$.
These values potentially come with an uncertainty or a systematic bias, coming from possible underestimation of other means of transportation such as cars; unaccounted patterns of local travel; or errors from an extrapolation procedure from historical data from other time periods.
In this section, we address the robustness of our inference procedure by testing how systemic and random deviations on $f_{ij}$ parameters affect the inferred parameter values.

\begin{figure}
    \includegraphics{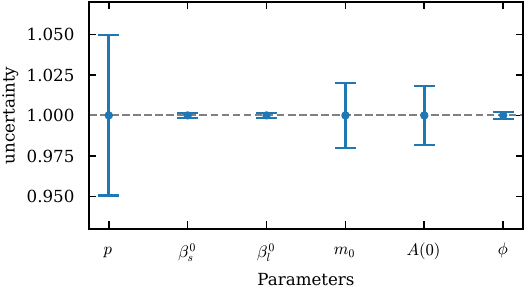}
    \caption{\label{fig:travel_noise} Uncertainty of parameters estimation measured in the units of relative standard deviation, under migration rates $f_{ij}$ with added Gaussian noise.
    The noise standard deviation is equal to 30\% of the original values of $f_{ij}$.
    Bars represent relative standard deviation -- standard deviation divided by average value -- computed over different noise realisations.}
\end{figure}

First, we test how random noise applied to the migration rates, impacts the inference procedure.
To this end, we add a zero-mean Gaussian noise to each value in Table \ref{tab:fij} (keeping the flow matrix symmetric) with standard deviation equal to 30\% of the original values.
Such level of noise variability is significant enough to take into account for a potentially high error on migration rates.
The updated values of $f_{ij}$ are subsequently used to infer all the parameters from the real case count data.
We repeat this procedure 5 times with different noisy realisations of $f_{ij}$, and report the relative uncertainty for each inferred parameter in Fig. \ref{fig:travel_noise}.
The bars show the relative standard deviation, computed as the ratio between standard deviation and parameter average value, over the values inferred originally.
For example, the fraction of asymptomatic cases $p$ have almost 5\% uncertainty, or in other words, its standard deviation in this experiment is equal to almost 5\% of its originally inferred value.
Note that all of the parameters have comparably low relative standard deviations, which indicates the robustness of our inference approach to a misspecification of migration rates $f_{ij}$.
Moreover, similarly to the results in section \ref{app:synth} and Table \ref{tab:all_results}, direct transmission rates $\beta_s$ and $\beta_l$ are inferred with the best precision, while parameter $p$ has the highest uncertainty.
This observation is in line with the principle discrepancy observed between the inference results from case-count and the genetic sequences, discussed in the main text.

\begin{figure}
    \includegraphics{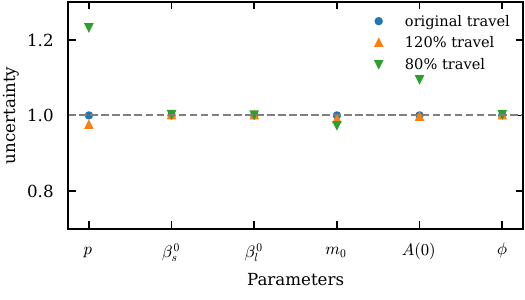}
    \caption{\label{fig:travel_bias} Values of inferred parameters divided by the results from Table \ref{tab:all_results}, for different values of travelling rates $f_{ij}$.
    Blue circles represent results obtained with the migration rates computed in section \ref{app:fij} (all ratios are equal to $1$), orange upper-triangles represent results obtained with travelling rates multiplied by $1.2$, green down-triangles represent results obtained with travelling rates multiplied by $0.8$.}
\end{figure}

Second, we analyse how re-scaling of migration rates (which mimics the bias in the estimation of the travel flows) changes the outcome of the inference procedure.
To this end, we multiply all values of $f_{ij}$ by a constant and infer parameters as in the original approach.
Fig. \ref{fig:travel_bias} shows how decreasing or increasing travelling rates by $20\%$ affects the values of inferred parameters.
In both cases the majority of inferred parameters remain robustly stable.
The most significantly affected ones are: the fraction of asymptomatic $p$ and the number of initially infected cases $A(0)$ (as a reminder, we note that all of the initial cases are assumed to be asymptomatic in our meta-population model).
Interestingly, decreased migration rates produced a stronger effect compared to increased ones.
Given that the number of travelling people has been likely under-estimated given that we did not include all possible transportation ways in our estimation procedure in section \ref{app:fij}, we conclude that the inference results obtained from the case counts are substantially robust to fluctuations and systemic bias in the migration rates used in our dynamic model.

\section{Impact of different cut-off cumulative horizons in the sampling procedure}
\label{app:sampling}
As described in the main text and in section \ref{app:samp_procedure}, in the absence of a recent preceding sampling date, we set the cut-off cumulative horizon to 14 days preceding the current sampling date. In reality it is difficult to set this parameter from first principles, and the only information available is the inverse of the recovery rate $1/\gamma = 7$ which has a meaning of an expected wait time to recovery.
In our simulations, we used the double of this period to account for possible variations in the recovery in our sampling model connecting the continuous model with the discrete predictions, which is to a certain extent an arbitrary modeling choice. It is reasonable that this cut-off cumulative horizon is larger than 7 days to allow for longer recovery times, but it should probably not be too large to avoid over-counting.
To assess an impact of the length of the cut-off cumulative horizon on the inference results, we repeat the whole inference procedure for smaller and larger values of the time interval,  7 and 21 days.

\begin{table}
\caption{\label{tab:sampling} Model parameters inferred from the case-count data with the sampling procedure using cut-off cumulative period of 7 and 21 days.}
\begin{ruledtabular}
\begin{tabular}{|l|cc|}
parameter & 7 days & 21 days \\
\hline
$p$ & 0.000 & 0.632 \\
$\beta_s$ & 0.158 & 0.160 \\
$\beta_l$ & 0.156 & 0.157 \\
$\phi$ & 5.75 & 5.89 \\
$m_0$ & 10.25 & 9.50 \\
$A(0)$ & 10.7 & 12.6 \\
\end{tabular}
\end{ruledtabular}
\end{table}

As shown in Table \ref{tab:sampling}, overall both analysed scenarios produce similar results to the ones presented in the main text.
Most of the parameters are very close to the inference using the 14 days sampling horizon scenario, except for the fraction of asymptomatic $p$.
In the case compatible with the length of the inverse recovery rate, this parameter predicts an unrealistic value $p=0$, while the long sampling horizon scenario leads to the value $p=0.632$, similar to what was obtained for two peaks in Appendix \ref{app:horizon}.
This suggests that as long as the cut-off cumulative period is larger than the inverse recovery rate in the sampling procedure, it seems to have a weak effect on the values of the inferred parameters, and illustrates once again the sensitivity of the fraction of asymptomatic infections $p$.

\section{Impact of the data length on the inferred parameter values}
\label{app:horizon}
From our analysis presented in the main text, we see that the fraction of asymptomatic infections $p$ inferred from genomic sequence data is significantly higher compared to the value estimated from the case count data. One of the main differences between the used data sets is their length: the used data on genomic sequences is focused on a single year (1993) of the outbreak, while the case-count data spans the period of 7 years, from 1992 to 1998.
In this section, we study the influence of time horizon on the results of the inference from the case counts, in order to test if the the data length could be the reason of the discrepancy in the estimation of the $p$ parameter. 
To this end, we repeat the inference procedure using two shorter case-count data sets.
In the first case we limit ourselves only to cases sampled between February 8th 1992 and August 1st 1993 (only one peak in Fig. \ref{fig:samples}), while in the second case, between February 8th 1992 and September 1st 1994 (only two peaks in Fig. \ref{fig:samples}).

\begin{table}
\caption{\label{tab:horizon} Model parameters inferred from the case count data restricted to the first peak or the first two peaks in Fig. \ref{fig:samples}.}
\begin{ruledtabular}
\begin{tabular}{|l|cc|}
parameter & one peak & two peaks \\
\hline
$p$ & 0.985 & 0.606 \\
$\beta_s$ & 0.248 & 0.164 \\
$\beta_l$ & 0.227 & 0.161 \\
$\phi$ & 3.19 & 5.96 \\
$m_0$ & 3.81 & 7.33 \\
$A(0)$ & $5 \cdot 10^{-6}$ & 11.9 \\
\end{tabular}
\end{ruledtabular}
\end{table}

The parameters inferred from these two parts of the data set are shown in Table \ref{tab:horizon}.
If we include only the data from the first peak, we get a high fraction of asymptomatic infections, which is much closer to the value obtained from the pathogen sequence data.
At the same time, we the inferred values of $\beta_s$ and $\beta_l$ turn out much higher compared to the results of inference from the full data set. Most importantly, the inferred number of initial cases $A(0)$ is smaller than one person, and the model with the inferred parameters is not able to reproduce more than one peak of the data, despite the seasonal form of $\beta_s(t)$ and $\beta_l(t)$ in Eq. (\ref{eq:beta}).
This periodic dependence accounting for seasonality can also probably be the source of significant discrepancies between the parameters obtained from the whole data-set and from the time horizon that only includes a single peak.

Using the data length that includes the first two peaks produces parameters that are much more consistent with the results obtained from inference using the full case-count data set.
Furthermore, this time the model is able to reproduce multiple peaks (and not only the first two).
The major difference is in the fraction of asymptomatic infections $p$, which is inferred at the level of $0.6$: in between the result obtained from the full case-count data ($p=0.32$), and the result inferred from the pathogen sequence data ($p=0.94$).
This results provides an additional evidence for the highest uncertainty associated with the recovery of the parameter $p$ compared to other model parameters. It may also indicate that the difference in the span of the data could partially contribute to the discrepancy between the inferred values of $p$ from case counts and from genomic sequence data.

\section{Implementing the cholera dynamic model for the phylogenetic analysis in PhyDyn}
\label{app:phylogenetic_analysis}
We use the PhyDyn library v.1.3.6 \cite{volz2018bayesian} in BEAST2 v.2.6.1 \cite{bouckaert2019beast} to estimate time-resolved phylogenies and epidemiological parameters from 186 sequences sampled between November 13, 1992 and March 31, 1993 across the three city regions (55 in Tartagal, 90 in Jujuy, and 41 in Oran). 

PhyDyn is a coalescent-based inference approach that simultaneously estimates both the pathogen phylogeny and epidemiological parameters for a defined set of ordinary differential equations (ODEs) from the pathogen sequence data. The PhyDyn framework decomposes the ODE equations into birth $F(t)$ and migration matrices $G(t)$ for infected hosts. For our three city cholera model in Eq. (\ref{eq:ode}), the matrices are: 

\begin{widetext}
\begin{align*}
    \text{demes} = \begin{Bmatrix}I_1 & A_1 & I_2 & A_2 & I_3 & A_3 \end{Bmatrix},
\end{align*}

\begin{align*}
    \mu(t)^T =
    \begin{pmatrix}
        \gamma I_1(t) & \gamma A_1(t) & \gamma I_2(t) & \gamma A_2(t) & \gamma A_3(t) & \gamma A_3(t)
    \end{pmatrix},
\end{align*}

\begin{align*}
    \small
    F(t) =
    \begin{pmatrix}
        (1-p) \beta(t)I_1(t) S_1(t) & p \beta(t) I_1(t) S_1(t) & 0 & 0 & 0 & 0 \\
        (1-p) \beta(t)A_1(t) S_1(t) & p \beta(t) A_1(t) S_1(t) & 0 & 0 & 0 & 0 \\
        0 & 0 & (1-p) \beta(t)I_2(t) S_2(t) & p \beta(t) I_2(t) S_2(t) & 0 & 0 \\
        0 & 0 & (1-p) \beta(t)A_2(t) S_2(t) & p \beta(t) A_2(t) S_2(t) & 0 & 0 \\
        0 & 0 & 0 & 0 & (1-p) \beta(t)I_3(t) S_3(t) & p \beta(t) I_3(t) S_3(t) \\
        0 & 0 & 0 & 0 & (1-p) \beta(t)A_3(t) S_3(t) & p \beta(t) A_3(t) S_3(t) \\
     \end{pmatrix},
\end{align*}

\begin{align*}
    G(t) =
    \begin{pmatrix}
        0 & 0 & 0 & 0 & 0 & 0\\
        0 & 0 & 0 & f_{21} \frac{A_1(t)}{S_1(t) + A_1(t) + R_1(t)} & 0 & f_{31} \frac{A_1(t)}{S_1(t) + A_1(t) + R_1(t)} \\
        0 & 0 & 0 & 0 & 0 & 0 \\
        0 & f_{12} \frac{A_2(t)}{S_2(t) + A_2(t) + R_2(t)} & 0 & 0 & 0 & f_{32} \frac{A_2(t)}{S_2(t) + A_2(t) + R_2(t)}  \\
        0 & 0 & 0 & 0 & 0 & 0 \\
        0 & f_{13} \frac{A_3(t)}{S_3(t) + A_3(t) + R_3(t)} & 0 & f_{23} \frac{A_3(t)}{S_3(t) + A_3(t) + R_3(t)} & 0 & 0 \\
     \end{pmatrix},
\end{align*}
\end{widetext}
where the infected hosts are the symptomatic $I_i$ and asymptomatic $A_i$ compartments of each city $i$.

The birth matrix $F(t)$ captures the birth rate of new infections in different population compartments, or demes, at time $t$.  In $F(t)$, the $(ij)$th element is the expected number of secondary infections in deme $i$ caused by a single infected individual in deme $j$. In our model, new infections come from both the asymptomatic and symptomatic individuals of the three cities. The migration matrix $G(t)$ tracks the flow of infectious individuals. In $G(t)$, the $(ij)$th element is the expected number of infectious individuals in deme $i$ who arrive in deme $j$.  As stated in the assumptions, we only account for travel of asymptomatic individuals. The time-dependent $\mu$ vector contains the recovery rates.  

The model demographic initial conditions were set to those described in the main text (See \emph{Materials and Methods}, section on the inference of epidemiological parameters from case count data); the seasonality and epidemiological parameters were set to those inferred in Table \ref{tab:all_results}); and the traveling values $f_{ij}$ were set to those in Table \ref{tab:fij}.  

We tested various combinations of molecular clock---the model of how quickly and with how much heterogeneity lineages accumulate genetic mutations---and substitution models, finding that the default HKY substitution model \cite{hasegawa1985dating} that allows for transition and transversion mutations to occur at different relative rates defined by the multiplicative factor $\kappa$ (lognormal prior for $\kappa$ with log mean = 1.0, standard deviation = 1.25) and a strict molecular clock produced the best mixing in a reasonable computational time. To set the prior on the clock rate, we first used treedater R package \cite{volz2017scalable} to estimate the evolutionary rate assuming a strict molecular clock, and then used the log of the estimated value as our prior mean (log mean = -11.5, standard deviation = 1.0). We assumed the proportion of invariant sites was 0, which in this case was known \textit{a priori} because only positions with at least one polymorphism were available in Genbank. All other priors were kept at their default values. We explored the possibility of sampling both the tree and model space jointly.  While we found this to be possible, it was computational inefficient (e.g., weeks of calendar time per million samples). Therefore, to simplify the inference, we fixed the tree topology to its maximum likelihood topology (presented in the Fig. \ref{fig:tree} in the main text) inferred using IQtree2 \cite{minh2020iq}.

\section{Pathogen sequence data: Inference sensitivity of epidemiological parameters to the number of fixed parameters}
\label{app:genomic_inference_sensitivity}

As described in the main text, we explore the sensitivity of estimates of $\beta_s$ and $\beta_l$ to different subsets of fixed epidemiological parameters.
One setting estimates both $\beta_s$ and $\beta_l$ (setting 1), and a more unconstrained setting estimates $\beta_s$,  $\beta_l$, and $p$ (setting 2). All remaining parameters were set to the maximum likelihood value estimated from the case count data (Table \ref{tab:all_results}). To account for the fact that the genetic sequence model only had one year of data we use a lognormal prior (log mean = -1.85, log standard deviation = 1.0) for both $\beta_s$ and $\beta_l$ to discourage transmission rates that were large enough to cause the entire population to become infected in a single year. For the setting 2, we used a uniform prior $U(0,1)$ for $p$.  

The posterior densities for the parameter estimates for both transmission rate parameters are shown in Table \ref{tab:genetic_results_main} in the main text.
When estimating both transmission terms in settings 1 and 2,  $\beta_s$ estimates are consistently higher than $\beta_l$, as also seen in the case count model inference. While the total number of infections is very similar under settings 1 and 2, the number of symptomatic infections is much smaller under setting 2 due to the high estimate of $p$. However, in setting 2 the 90th percentile of the peak prevalence is about 8, which is close to the case-count predicted values, that is, the prediction for the prevalence of symptomatic cases under setting 2 is lower than in the case-count model. However, we find that the value is still well within the range of values given the confidence intervals for the case-count model.

\begin{figure*}
    \centering
    \subfloat[Tartagal]{\includegraphics{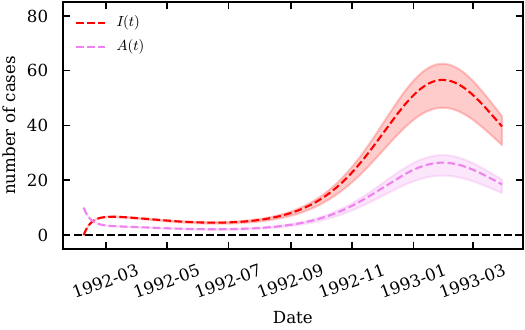}}\\
    \subfloat[San Ramón de la Nueva Orán (Oran)]{\includegraphics{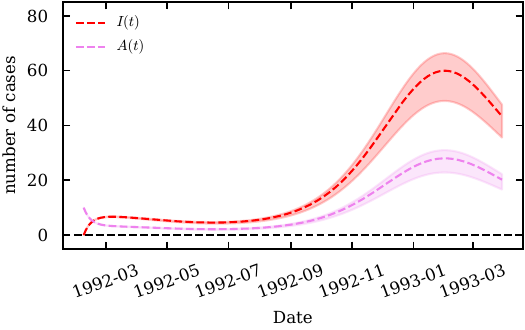}}\\
    \subfloat[San Salvador de Jujuy (Jujuy)]{\includegraphics{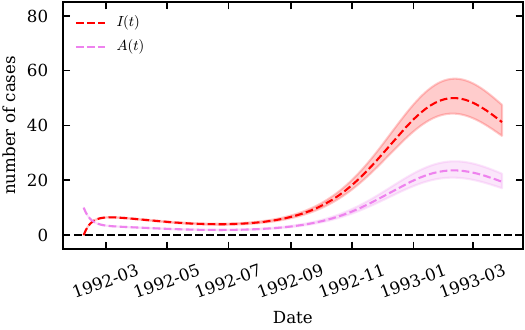}}\\
    \caption{\label{fig:traj2} Model predicted prevalence of cholera in three cities based on the genetic sequence data (setting 1, $p$ fixed from case-count data inference).
    Red and purple dashed lines represent the medians of expected number of symptomatic and asymptomatic infected cases accordingly as a function of time.
    Lighter bands around the dashed lines represent the space between 10th and 90th percentiles.}
\end{figure*}

Posterior trajectories for setting 1 are shown in Fig. \ref{fig:traj2}.
The peak prevalence is about three times higher using the point estimates for $\beta$ from the genetic sequence data.
However, overall the estimates of $\beta$ from the genetic sequence data was very similar to those obtained by the case-count model.

\begin{figure*}
    \centering
    \subfloat[Tartagal]{\includegraphics{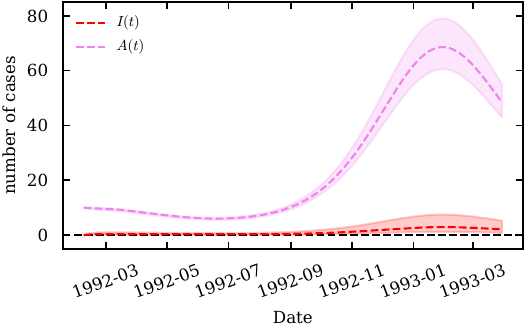}}\\
    \subfloat[San Ramón de la Nueva Orán (Oran)]{\includegraphics{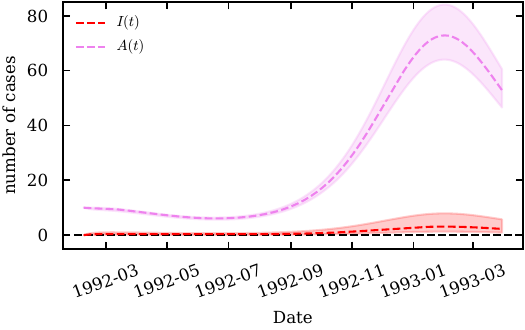}}\\
    \subfloat[San Salvador de Jujuy (Jujuy)]{\includegraphics{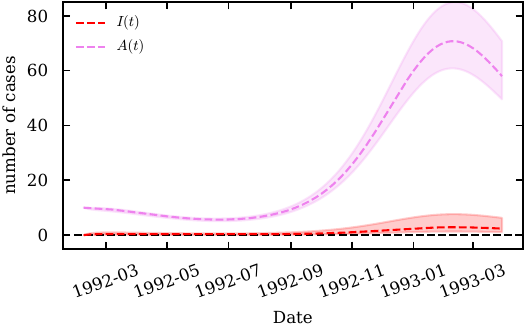}}\\
    \caption{\label{fig:traj3} Model-predicted prevalence of cholera in three cities based on the genetic sequence data (setting 2, $p$ is a free parameter inferred from the sequence data).
    Red and purple dashed lines represent the medians of expected number of symptomatic and asymptomatic infected cases accordingly as a function of time.
    Lighter bands around the dashed lines represent the space between 10th and 90th percentiles.}
\end{figure*}

The point estimate of $p$ (0.94, 95\% HPDI 0.85, 0.99) under setting 2 was quite different from the case-count model, which once again illustrates the variation of the inferred value of this parameter under different inference settings.
The posterior trajectories for setting 2 are shown in Fig. \ref{fig:traj3}.

\end{document}